\documentclass[twocolumn]{pasj02} 
\jyear{2025}
\Received{}%{yyyy/mm/dd}
\Accepted{}%{yyyy/mm/dd}
\begin{document} 

\title{CO Multi-line Imaging of Nearby Galaxies (COMING). XI. Azimuthally averaged star formation rate and stellar mass relation with molecular gas amount}

%%% begin:list of authors
% Do NOT capitalize all letters in "textsc".
\author{
 Ayumi \textsc{Kajikawa},\altaffilmark{1} 
 Kazuo \textsc{Sorai},\altaffilmark{1,2}\altemailmark\orcid{0000-0003-1420-4293} \email{sorai@phys.sci.hokudai.ac.jp} 
 Kana \textsc{Morokuma-Matsui},\altaffilmark{3}\orcid{0000-0003-3932-0952} 
 Tsutomu T. \textsc{Takeuchi},\altaffilmark{4,5}\orcid{0000-0001-8416-7673} 
 Dragan \textsc{Salak},\altaffilmark{1,6}\orcid{0000-0002-3848-1757} 
 Nario \textsc{Kuno},\altaffilmark{7,8}\orcid{0000-0002-1234-8229} 
 Kazuyuki \textsc{Muraoka},\altaffilmark{9}\orcid{0000-0002-3373-6538} 
 Yusuke \textsc{Miyamoto},\altaffilmark{10,11}\orcid{0000-0002-7616-7427} 
 Hiroyuki \textsc{Kaneko},\altaffilmark{10,12}\orcid{0000-0002-2699-4862} 
 Yoshiyuki \textsc{Yajima},\altaffilmark{1}\orcid{0000-0002-5476-9101} 
 Atsushi \textsc{Yasuda},\altaffilmark{7}\orcid{0000-0003-2076-8685} 
 Takahiro \textsc{Tanaka},\altaffilmark{7} 
 Kiyoaki Christopher \textsc{Omori},\altaffilmark{4}\orcid{0000-0002-8432-6870} 
 Kazuki \textsc{Shimizu},\altaffilmark{1} 
 Suphakorn \textsc{Suphapolthaworn},\altaffilmark{1}\orcid{0009-0007-4497-3954} 
 and 
 Kyoko \textsc{Hama}\altaffilmark{1} 
}
\altaffiltext{1}{Department of Cosmosciences, Graduate School of Science, Hokkaido University, Kita 10, Nishi 8, Kita-ku, Sapporo, Hokkaido 060-0810, Japan}
\altaffiltext{2}{Department of Physics, Faculty of Science, Hokkaido University, Kita 10, Nishi 8, Kita-ku, Sapporo, Hokkaido 060-0810, Japan}
\altaffiltext{3}{Institute of Astronomy, Graduate School of Science, The University of Tokyo, 2-21-1, Osawa, Mitaka, Tokyo 181-0015, Japan}
\altaffiltext{4}{Division of Particle and Astronautical Science, Nagoya University, Furo-cho, Chikusa-ku, Nagoya, Aichi 464-8602, Japan}
\altaffiltext{5}{The Research Center for Statistical Machine Learning, the Institute of Statistical Mathematics, 10-3 Midori-cho, Tachikawa, Tokyo 190-8562, Japan}
\altaffiltext{6}{Institute for the Advancement of Higher Education, Hokkaido University, Kita 17, Nishi 8, Kita-ku, Sapporo, Hokkaido 060-0817, Japan}
\altaffiltext{7}{Graduate School of Pure and Applied Sciences, University of Tsukuba, 1-1-1 Tennodai, Tsukuba, Ibaraki 305-8571, Japan}
\altaffiltext{8}{Tomonaga Center for the History of the Universe, University of Tsukuba, 1-1-1 Tennodai, Tsukuba, Ibaraki 305-8571, Japan}
\altaffiltext{9}{Graduate School of Science, Osaka Metropolitan University, 1-1 Gakuen, Sakai, Osaka 599-8531, Japan}
\altaffiltext{10}{National Astronomical Observatory of Japan, 2-21-1, Osawa, Mitaka, Tokyo 181-8588, Japan}
\altaffiltext{11}{Department of Electrical, Electronic and Computer Engineering, Fukui University of Technology, 3-6-1, Gakuen, Fukui 910-8505, Japan}
\altaffiltext{12}{Institute of Science and Technology, Niigata University, 8050 Ikarashi 2-no-cho, Nishi-ku, Niigata 950-2181, Japan}

%%% end:list of authors

%% !!! Select 3 to 5 words from PASJ's key words !!! 
%% List of Key Words: https://academic.oup.com/pasj/pages/Pasj_Keywords 
%% "\KeyWords{ }" always has to be placed before ``\maketitle'' 

\maketitle

\begin{abstract}
This study investigated the relation between the surface density of star formation rate (SFR) ($\Sigma_{\mathrm{SFR}}$), stellar mass ($\Sigma_{M_{\ast}}$), and molecular gas mass ($\Sigma_{M_\mathrm{mol}}$) on nearly 1\,kpc scales averaged over concentric tilted rings using the \atom{C}{}{12}\atom{O}{}{} $J=1-0$ mapping data of 92 nearby galaxies obtained in the CO Multi-line Imaging of Nearby Galaxies (COMING) project. 
We categorized these galaxies into three groups based on the deviation of each global SFR from the star-forming main sequence (MS), $\Delta$MS: upper MS (UMS), MS, and lower MS (LMS). 
UMS galaxies tend to be less massive or barred spiral galaxies, exhibiting molecular gas fraction ($f_{\mathrm{gas}}$) comparable to those of MS galaxies but higher star formation efficiency (SFE). 
In contrast, the LMS galaxies tend to be massive or active galaxies hosting an active galactic nucleus (AGN). 
Their $f_{\mathrm{gas}}$ values are lower than those of MS galaxies, and their SFEs are slightly lower or comparable to those of MS galaxies in the inner region. 
These trends indicate that enhanced SFE contributes to higher $\Delta$MS values, whereas reduced $f_{\mathrm{gas}}$ results in lower $\Delta$MS values. 
The less prominent bulge or the presence of a bar structure in UMS galaxies induces disk-wide star formation, consequently increasing the SFE. 
In LMS galaxies, the molecular gas is exhausted, and their star formation activity is low. 
Environmental effects, such as tidal gas stripping, may also reduce gas supply from the outer regions. 
Furthermore, our sample galaxies show that both the specific star formation rate (sSFR) and $f_{\mathrm{gas}}$ decrease in the central region in LMS galaxies but did not change in the same region in UMS galaxies. 
These results seem to support the inside--out quenching of star formation although the dominant cause of depletion remains uncertain.
\end{abstract}

%\pagewiselinenumbers 

\section{Introduction}

%\noindent IMPORTANT NOTICE\\
%1. Manuscript for submission must be in the same format as a published papers. \\
%2. Line numbers should be added to the manuscript. \\
%3. Do NOT use ``\verb|\def|, \verb|\renewcommand|''.\\
%4. Do NOT redefine commands provided by pasj02.cls.  

A strong relationship between the global\footnote{Throughout the paper, we imply the integration over a galaxy as ``global.''} star formation rate (SFR) and stellar mass ($M_{\ast}$) of galaxies has been widely known as the star-forming main sequence (SFMS) (e.g., \cite{Noeske07, Elbaz07, Daddi07}). 
Star formation is ongoing in galaxies located on and above the SFMS, whereas galaxies whose star-formation activity is quenched lie below this sequence on the scatter plot between the SFR and $M_{\ast}$. 
As star formation is an essential process that accounts for galaxy evolution, the location of galaxies on the SFR--$M_{\ast}$ plot can be a clue to explore this evolution. 
Although the general trend {\it along} the SFMS is well-known, recent studies have focused on the {\it scattering} around it to determine the physics driving these deviations (e.g., \cite{Lin19}; \cite{Ellison20b}; \cite{Colombo25b}).

The surface density of the SFR ($\Sigma_{\mathrm{SFR}}$) and stellar mass ($\Sigma_{M_{\ast}}$) on the nearly 1\,kpc scale also have a similar strong correlation, termed as the spatially resolved SFMS (rSFMS). 
Large integral field spectroscopic (IFS) surveys have revealed this relationship in galaxies with low redshifts. 
\citet{Cano-Diaz16} used the Calar Alto Legacy Integral Field Area (CALIFA) survey (\cite{Sanchez12}) data and found rSFMS on 0.5--1.5\,kpc scales. 
\citet{Hsieh17} used the Mapping Nearby Galaxies at the Apache Point Observatory (MaNGA) (\cite{Bundy15}) and found the rSFMS on the kpc scale. 
These results show that this subgalactic relation is more fundamental to understanding galaxy evolution and that the global properties of galaxies originate from the local star formation mechanism. 
Therefore, further investigation of rSFMS provides a key to understanding the processes governing star formation on a global scale.

Recent studies have shown that galaxies with different offsets from the global SFMS show different trends in the spatially resolved $\Sigma_{\mathrm{SFR}}$ -- $\Sigma_{M_{\ast}}$ relation. 
\citet{Ellison18} investigated the spatially resolved $\Sigma_{\mathrm{SFR}}$ -- $\Sigma_{M_{\ast}}$ relation and the dependence of enhanced/suppressed star formation for galaxies that lie above/below the global SFMS using the MaNGA data. 
They found that galaxies above the global SFMS have higher $\Sigma_{\mathrm{SFR}}$ values in the central region, whereas those below the global SFMS, i.e., passive galaxies, have lower $\Sigma_{\mathrm{SFR}}$ values over the galactic disks, with the most significant decrease in the central region. 
Abdurro'uf \& Akiyama (2017, 2018) also found a flattening $\Sigma_{\mathrm{SFR}}$ at high $\Sigma_{M_{\ast}}$ in the subgalactic scale relation for galaxies that lie below the global SFMS. 
These trends indicate a more rapid decrease in the specific star formation rate ($\mathrm{sSFR} \equiv \mathrm{SFR} / M_{\ast}$) in the central region than the outskirts for passive galaxies, and agree with the inside--out quenching scenario (e.g., Tacchella \etal 2015, 2018; \cite{GonzalezDelgado16}; \cite{Belfiore18}; \cite{Matharu22}; \cite{Wylezalek22}). 
Although several physical processes have been proposed as the quenching mechanism, the physical mechanism responsible for the inside--out quenching remains unclear.

Star formation activity is expected to depend on both the amount of molecular gas and the conversion efficiency from molecular gas to stars. 
Previous studies have shown that galaxies above the global SFMS have a higher global molecular gas mass fraction ($f_{\mathrm{gas}} \equiv \mathrm{global\ molecular\ gas\ mass}\ M_{\mathrm{mol}} / M_{\ast}$, e.g., \cite{Scoville16, Lin17}), global star formation efficiency ($\mathrm{SFE} \equiv \mathrm{SFR} / M_{\mathrm{mol}}$, e.g., \cite{Saintonge12, Sargent14, Genzel15, Silverman15}, 2018; \cite{Tacconi18}), or a combination of both (\cite{Saintonge12}, 2016; \cite{Combes13, Scoville16, Lin17, Elbaz18}). 
Additionally, subsequent studies have explored the relationships among molecular gas, stellar mass, and SFR on a subgalactic scale. 
Based on the ALMA--MaNGA QUEnching and STar formation (ALMaQUEST) survey (\cite{Lin20}), \citet{Lin19} observed that rSFMS results from the combination of two relationships: 
(1) molecular-gas main sequence (MGMS, molecular gas mass surface density $\Sigma_{M_{\mathrm{mol}}}$ -- $\Sigma_{M_{\ast}}$), which regulates the amount of molecular gas in the interstellar medium; and 
(2) the Kennicutt--Schmidt (K--S) relation ($\Sigma_{M_{\mathrm{mol}}}$ -- $\Sigma_{\mathrm{SFR}}$, e.g., \cite{Schmidt59}; \cite{Kennicutt89}; \cite{Kennicutt12}), which describes the conversion of molecular gas into stars. 
\citet{Ellison20b} used maps with a kpc-scale resolution of 46 galaxies in the ALMaQUEST to investigate mechanisms regulating star formation. 
They proposed that the scattering of the rSFMS is primarily driven by changes in the local SFE, with a secondary dependence on the local $f_{\mathrm{gas}}$. 
Such kpc-scale discussions have also been provided by the Extragalactic Database for Galaxy Evolution Calar Alto Legacy Integral Field Area (EDGE-CALIFA) survey (\cite{Bolatto17}). 
Recently, \citet{Ellison24} demonstrated the importance of dynamical equilibrium pressure in regulating the SFR.

Galactic structures and environmental effects are also expected to induce variations in the SFE from the molecular gas in galaxies. 
\citet{Ellison20a} used 12 central starburst galaxies above the global SFMS and found that the starburst in the central region is primarily characterized by an enhanced SFE. 
The centrally enhanced SFE is affected by morphological features, such as bars and/or mergers (e.g., \cite{Bolatto17, Utomo17, Chown19}). 
However, few studies have investigated rSFMS combined with kpc-scale spatially resolved molecular gas information.

In this study, we used spatially resolved molecular gas information obtained from the CO Multi-line Imaging of Nearby Galaxies (COMING) project (\cite{Sorai19}). 
COMING observed 147 nearby galaxies with the 45\,m radio telescope of the Nobeyama Radio Observatory and obtained molecular gas images with a spatial resolution of $\sim 1$\ kpc in the \atom{C}{}{12}\atom{O}{}{} $J=1-0$ line. 
We examined the relationships among the surface densities of the SFR, stellar mass, and molecular gas at nearly 1\,kpc scale for 92 COMING galaxies. 
In addition, we investigated the relationship between variations in star formation activity, $f_{\mathrm{gas}}$, and SFE.

The remainder of this paper is organized as follows. 
In section \ref{sec:sample_data}, we describe the selection of sample galaxies and the data utilized to calculate $\Sigma_{M_{\ast}}$, $\Sigma_{\mathrm{SFR}}$, and $\Sigma_{M_{\mathrm{mol}}}$. 
The results are described in section \ref{sec:results}. 
We investigate the relation between global SFR and $M_{\ast}$ in subsection \ref{subsec:globalSFMS} and show the results of the relation between $\Sigma_{M_{\ast}}$, $\Sigma_{\mathrm{SFR}}$, and $\Sigma_{M_{\mathrm{mol}}}$ in subsection \ref{subsec:azimuthalavr}. 
In addition, we investigate the difference between the $\Sigma_{\mathrm{SFR}}$ and $\Sigma_{M_{\ast}}$ in subsection \ref{subsec:radialprof}. 
We discuss our results in section \ref{sec:discussion}, and summarize this research in section \ref{sec:conclusions}.

\section{Sample and data}\label{sec:sample_data}

In this section, we describe the selection of sample galaxies and the data used for deriving physical quantities.

\subsection{Sample}\label{subsec:sample}

As this study aimed to relate rSFMS with molecular gas content, we selected galaxies out of those observed in the COMING project with the following four criteria: 
\begin{enumerate}
\item Galaxies identified as spirals.
\item Galaxies not identified as edge-on.
\item Galaxies that are not in close pairs.
\item Galaxies with $\Sigma_{\mathrm{SFR}}$ measurements available. 
\end{enumerate}
The first criterion is set to compare galaxies without significantly different morphologies. 
The second is required to radially resolve each galaxy. 
The largest inclination angle of the samples was $75^{\circ}.8$. 
The third is set to exclude the effect of the interaction of galaxies on star formation and avoid the complexity of overlapped galaxy regions. 
The excluded galaxies are those listed as interacting galaxies in \citet{Sorai19}. 
Most of these are listed in the catalogue of interacting galaxies (\cite{Vorontsov-Velyaminov01}), 
while the remaining two pairs have also classified as interacting systems in previous studies: 
NGC\,4298/NGC\,4302 (e.g., \cite{Davis83}); NGC\,4383/UGC\,7504 (\cite{Koopmann04}).
The final sample size is 92, which is twice the number of galaxies investigated in the ALMaQUEST project. 
The following parameters of the sample galaxies are listed in table \ref{tab:galpars}: 
galaxy name, distance, position angle of the major axis of the galaxy disk (PA), inclination of the galaxy disk ($i$), $B$-band isophotal radius at 25\,mag\ arcsec$^{-2}$ ($R_{25}$), morphology (whether the galaxy has a bar or not), nuclear activity, nuclear star formation activity, environment, total stellar mass within $R_{25}$ ($M_{\ast}$), total SFR within $R_{25}$, and deviation from the SFMS ($\Delta$MS). 
The details of the $\Delta$MS are described in section \ref{sec:results}.

\subsection{Data}\label{subsec:data}

We used the two-dimensional surface density data of the molecular gas mass ($\Sigma_{M_{\mathrm{mol}}}$), stellar mass ($\Sigma_{M_{\ast}}$), and SFR ($\Sigma_{\mathrm{SFR}}$) of the target galaxies obtained by the COMING project. 
Herein, we briefly describe the derivation of these physical quantities from the data. 
A detailed description of the former two is presented in the COMING overview paper (\cite{Sorai19}), and the latter will be presented in a forthcoming paper (Takeuchi \etal in prep.).

$\Sigma_{M_{\mathrm{mol}}}$ is calculated from \atom{C}{}{12}\atom{O}{}{} $J=1-0$ integrated intensity $I_{^{12}\mathrm{CO}}$ maps obtained in the COMING project at an angular resolution of $\timeform{17''}$ with a $\timeform{6''}$ pixel size using the following formula: 
\begin{eqnarray}
\left( \frac{\Sigma_{M_{\mathrm{mol}}}}{M_{\solar}\ \mathrm{kpc}^{-2}} \right) &=& 3.20 \times 10^{6} \times 1.36\ \cos i \left( \frac{I_{^{12}\mathrm{CO}}}{\mathrm{K\ km\ s}^{-1}} \right) \nonumber \\
 &\times& \left[ \frac{X_{\mathrm{CO}}}{2.0 \times 10^{20}\ \mathrm{cm}^{-2}\ \left(\mathrm{K\ km\ s}^{-1}\right)^{-1}} \right]
\end{eqnarray}
We assume that the CO-to-H$_{2}$ conversion factor $X_{\mathrm{CO}}$ is constant for all the sample galaxies; $X_{\mathrm{CO}} = 2.0 \times 10^{20}\,\mathrm{cm}^{-2}\,\left(\mathrm{K\ km\ s}^{-1}\right)^{-1}$ (\cite{Bolatto13}). 
A factor of 1.36 is multiplied to include the contribution of He and heavier elements to $\Sigma_{M_{\mathrm{mol}}}$. 
The other factor $3.20 \times 10^{6}$ represents the coefficient for the conversion of units from the CO column density of cm$^{-2}$ to the surface density of the molecular gas of $M_{\solar}$\ kpc$^{-2}$.

$\Sigma_{M_{\ast}}$ is calculated from Wide-Field Infrared Survey Explorer (WISE) (\cite{Wright10}) 3.4\,$\mu$m band data. 
The pixel values of the stellar mass are derived from 3.4\,$\mu$m luminosity using the following formula (\cite{Wen13}): 
\begin{eqnarray}
\log_{10} \left( \frac{M_{\ast}}{M_{\solar}} \right) &=& (1.033 \pm 0.001) \log_{10} \left( \frac{\nu L_{\nu} (3.4\,\mu \mathrm{m})}{L_{\solar}} \right) \nonumber \\
&+&(0.679 \pm 0.002), 
\label{eq:calcMstar}
\end{eqnarray}
where $\nu$ is the effective wavelength in the band, and $L_{\nu} (3.4\,\mu \mathrm{m})$ is 3.4\,$\mu$m luminosity. 
Finally, the $\Sigma_{M_{\ast}}$ map is obtained by converting the stellar mass to surface density in each pixel. 
The angular resolution and grid size of this map are matched to the $\Sigma_{M_{\mathrm{mol}}}$ map of $\timeform{17''}$ and $\timeform{6''}$, respectively.

$\Sigma_{\mathrm{SFR}}$ is calculated from a combination of Galaxy Evolution Explorer (GALEX) (\cite{Martin05}) far-ultraviolet (FUV) and WISE 22\,$\mu$m intensities using the following formula (\cite{Casasola17}): 
\begin{eqnarray}
\left[ \frac{\Sigma_{\mathrm{SFR}}}{M_{\solar}\ \mathrm{yr}^{-1}\ \mathrm{kpc}^{-2}} \right] &=& \left[ 3.2 \times 10^{-3} \left( \frac{I_{22\,\mu\mathrm{m}}}{\mathrm{MJy\ sr}^{-1}} \right) \right. \nonumber \\
&+&\left. 8.1 \times 10^{-2} \left( \frac{I_\mathrm{FUV}}{\mathrm{MJy\ sr}^{-1}} \right)  \right] \nonumber \\
&&\times \cos i \times 1.59, 
\label{eq:calcSFR}
\end{eqnarray}
where $I_\mathrm{FUV}$ and $I_{22\,\mu\mathrm{m}}$ are FUV and 22\,$\mu$m intensities, respectively. 
The angular resolution and grid size of these maps are also matched to the $\Sigma_{M_{\mathrm{mol}}}$ map.

Using the maps of $\Sigma_{M_{\ast}}$, $\Sigma_{\mathrm{SFR}}$, and $\Sigma_{M_{\mathrm{mol}}}$, we calculated their azimuthally averaged values. 
We determined concentric tilted rings with $0.1 R_{25}$ width for each galaxy based on its PA and $i$ listed in table \ref{tab:galpars}. 
Here each $\timeform{6''} \times \timeform{6''}$ pixel was divided into $5 \times 5$ pixels to assign equally for every tilted ring. 
The pixels were only used when the $\Sigma_{M_{\ast}}$, $\Sigma_{\mathrm{SFR}}$, and $\Sigma_{M_{\mathrm{mol}}}$ have signal-to-noise ratio higher than or equal to five.

\section{Results}\label{sec:results}

\subsection{Global star formation main sequence}\label{subsec:globalSFMS}

\begin{figure}
 \begin{center}
  \includegraphics[width=8cm]{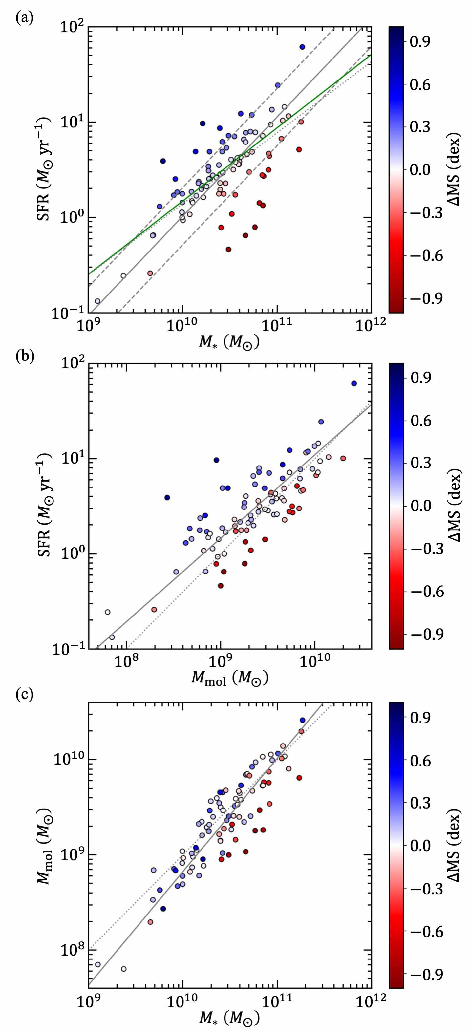}
 \end{center}
\caption{(a) The global SFR and stellar mass relation for 92 sample galaxies. 
The number of data points look smaller because of overlapping. 
The color corresponds to the offset from the star formation main sequence in our samples ($\Delta$MS). 
The gray solid line indicates the ordinary least-squares (OLS) bisector fitting for all data points. 
The gray dashed lines represent $\pm0.3$\,dex in SFR from the solid line. 
The gray dotted line represents the OLS fitting for the data. 
The best MS fit for star-forming galaxies observed at $z \sim 0$ with the Sloan Digital Sky Survey (SDSS) (\cite{Elbaz07}) is plotted as the green line. 
(b) Same as (a), except for the global SFR and molecular gas mass relation. 
The gray solid line indicates the OLS bisector fitting for all data points. 
The gray dotted line represents a constant star formation efficiency of $\mathrm{SFR} / M_{\mathrm{mol}} = 10^{-9}\ \mathrm{yr}^{-1}$. 
(c) Same as (b), except for the global molecular gas mass and stellar mass relation. 
The gray dotted line represents a constant molecular gas mass fraction of $M_{\mathrm{mol}} / M_{\ast} = 10^{-1}$. 
{Alt text: The upper graph shows that the global SFR is correlated with the total stellar mass, although scatter exists. 
Our adopted OLS bisector fitting line is slightly steeper than the line obtained in a previous study. 
The middle graph shows that the global SFR is correlated with the total molecular gas mass. 
Plots with higher delta main sequences tend to have higher star formation rates at the fixed molecular gas mass. 
The lower graph shows that the total molecular gas mass correlated well with the total stellar mass. 
The plots with higher delta main sequence tend to have higher molecular gas mass at the fixed stellar mass.}}
\label{fig:globalSFR_Mstar_Mmol}
\end{figure}

Figure \ref{fig:globalSFR_Mstar_Mmol}a shows the relation between the global SFR and $M_{\ast}$ for the 92 sample galaxies. 
The global SFR and $M_{\ast}$ values were estimated by summing up the values within $R_{25}$. 
This figure illustrates a positive correlation, as reported in previous studies. 
We determined the main sequence for the samples using least-squares fitting: 
\begin{equation}
\log_{10} \mathrm{SFR} = \alpha \log_{10} M_{\ast} + \beta
\end{equation}
indicated by the solid gray line in figure \ref{fig:globalSFR_Mstar_Mmol}a. 
Herein, we adopted an ordinary least-squares (OLS) bisector (\cite{Isobe90}) because, in the relation between SFR and $M_{\ast}$, one quantity should not be considered an independent variable and the other a dependent variable. 
The optimized parameters of this line are $\alpha = 1.04 \pm 0.06$ and $\beta = -10.4 \pm 0.6$. 
The dashed lines represent $\pm0.3$\,dex offset from the solid line along the vertical axis. 
For comparison with previous studies, the OLS fitting is represented as a gray dotted line in figure \ref{fig:globalSFR_Mstar_Mmol}a. 
Here $\alpha = 0.75 \pm 0.08$ and $\beta = -7.3 \pm 0.8$. 
This second method is similar to the best MS fit for star-forming galaxies observed at $z \sim 0$ with the Sloan Digital Sky Survey (SDSS) (\cite{Elbaz07}) plotted for comparison as a green line in figure \ref{fig:globalSFR_Mstar_Mmol}a. 
Although several studies on the SFMS have been conducted with numerous galaxies (e.g., \cite{Speagle14}), we used our MS for the 92 samples, as the relative difference of the SFR needed to be evaluated only at fixed $M_{\ast}$. 
We confirmed that the following results were not qualitatively affected, even when OLS fitting was used.

% figure 1: global SFR vs. M*, SFR vs. Mmol, and Mmol vs. M*

We categorized the 92 samples into three groups based on the deviation from the main sequence ($\Delta$MS), as this deviation has been considered to characterize the galaxy properties of star formation (e.g., \cite{Ellison20b}). 
$\Delta$MS is expressed as  
\begin{equation}
\Delta \mathrm{MS} (\mathrm{dex}) = \log_{10} \mathrm{SFR}(M_{\ast}) - \log_{10} \mathrm{SFR}_{\mathrm{fit}}(M_{\ast})
\end{equation}
where $\mathrm{SFR}_{\mathrm{fit}}$ is the global SFR when a global stellar mass $M_{\ast}$ is substituted into the equation of the least-squares power-law fit for all the data points. 
The sample galaxies are categorized based on their $\Delta$MS as follows: $\Delta \mathrm{MS} > 0.3$ (upper main sequence; UMS), $-0.3 \leq \Delta \mathrm{MS} \leq 0.3$ (main sequence; MS), and $\Delta \mathrm{MS} < -0.3$ (lower main sequence; LMS). 
A deviation value of 0.3\,dex was adopted, following \citet{Abdurrouf18}. 
The UMS, MS, and LMS contained 15, 63, and 14 galaxies, respectively.

The relation between the global SFR and $M_{\mathrm{mol}}$, and that between the global $M_{\mathrm{mol}}$ and $M_{\ast}$ for the 92 sample galaxies are also shown in figure \ref{fig:globalSFR_Mstar_Mmol}b and \ref{fig:globalSFR_Mstar_Mmol}c, respectively. 
We can observe the well-known K--S relation in figure \ref{fig:globalSFR_Mstar_Mmol}b; 
however, UMS galaxies show higher SFE than LMS galaxies in general. 
The global $f_{\mathrm{gas}}$ indicates the tightest relation compared with the global SFR--$M_{\ast}$ relation and global SFR--$M_{\mathrm{mol}}$. 
UMS galaxies tend to have higher $f_{\mathrm{gas}}$ than LMS galaxies.

\subsection{Azimuthally averaged relation among $\Sigma_{M_{\ast}}$, $\Sigma_{\mathrm{SFR}}$, and $\Sigma_{M_{\mathrm{mol}}}$}\label{subsec:azimuthalavr}

We investigated the relation of $\Sigma_{\mathrm{SFR}}$ -- $\Sigma_{M_{\ast}}$, $\Sigma_{\mathrm{SFR}}$ -- $\Sigma_{M_{\mathrm{mol}}}$, and $\Sigma_{M_{\mathrm{mol}}}$ -- $\Sigma_{M_{\ast}}$ planes, as reported in previous studies (\cite{Ellison20b, Morselli20}), which have suggested that these surface densities are strongly correlated with one another and form a three-dimensional linear relation (\cite{Lin19, Enia20}). 
The surface densities were averaged over concentric tilted rings with widths of $0.1\,R_{25}$, as described in subsection \ref{subsec:data}, to represent the radial variations in star formation in galaxies.

\begin{figure}
 \begin{center}
  \includegraphics[width=8cm]{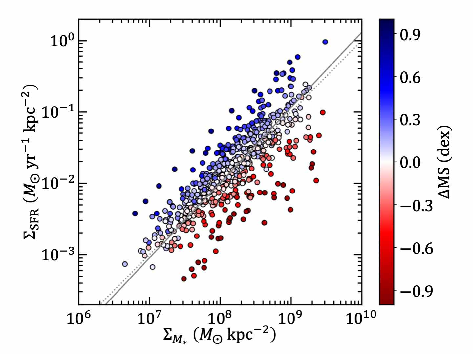}
 \end{center}
\caption{Azimuthally averaged SFR and stellar mass surface density relation for the 92 sample galaxies. 
The color corresponds to the offset from the star formation main sequence in our samples ($\Delta$MS). 
The gray solid line represents the OLS bisector fitting for all data points. 
The gray dotted line represents a constant specific star formation rate of $\Sigma_{\mathrm{SFR}} / \Sigma_{M_{\ast}} = 10^{-10}\ \mathrm{yr}^{-1}$. 
{Alt text: This graph show that the SFR surface density correlates well with the stellar mass surface density. 
Plots with higher delta main sequence tend to have higher SFR surface density at the fixed stellar mass surface density.}}
\label{fig:resolvedSFR_Mstar}
\end{figure}

Figure \ref{fig:resolvedSFR_Mstar} shows a correlation between $\Sigma_{\mathrm{SFR}}$ and $\Sigma_{M_{\ast}}$. 
Each point is color-coded by the $\Delta$MS derived from the global MS relationship in each galaxy. 
The solid line represents the OLS bisector power-law fit for all data points, where the slope and intercept are $1.05 \pm 0.03$ and $-10.4 \pm 0.2$, respectively. 
The dotted line represents a constant $\mathrm{sSFR} \equiv \Sigma_{\mathrm{SFR}} / \Sigma_{M_{\ast}} = 10^{-10}\ \mathrm{yr}^{-1}$. 
The azimuthally averaged $\Sigma_{\mathrm{SFR}}$ and $\Sigma_{M_{\ast}}$ also have a good correlation, as well as the global SFR and $M_{\ast}$ relation. 
This good correlation is consistent with those reported in previous studies on spatially resolved SFMS (e.g., \cite{Wuyts13, Hsieh17, Ellison20a}). 
Moreover, we found that the azimuthally averaged $\Sigma_{\mathrm{SFR}}$ and $\Sigma_{M_{\ast}}$ relation had a gradient of $\Delta$MS, that is, high sSFR values were preferentially registered in UMS galaxies. 
Additionally, $\Sigma_{\mathrm{SFR}}$ at a high $\Sigma_{M_{\ast}}$ region behaves differently based on $\Delta$MS. 
UMS galaxies show that $\Sigma_{\mathrm{SFR}}$ gradually increases with $\Sigma_{M_{\ast}}$ following the SFMS, whereas the $\Sigma_{\mathrm{SFR}}$ of the LMS galaxies tends to be flattened at high $\Sigma_{M_{\ast}}$.

% figure 2: azimuthally averaged surface density of SFR vs. that of M*

\begin{figure}
 \begin{center}
  \includegraphics[width=8cm]{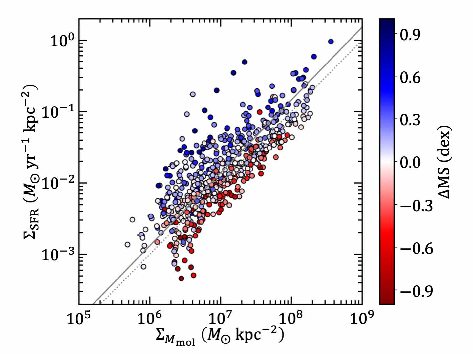}
 \end{center}
\caption{Same as figure \ref{fig:resolvedSFR_Mstar}, except for the relation between averaged SFR and molecular gas mass surface density. 
The gray solid line represents the OLS bisector fitting for all data points. 
The gray dotted line represents a constant star formation efficiency of $\Sigma_{\mathrm{SFR}} / \Sigma_{M_{\mathrm{mol}}} = 10^{-9}\ \mathrm{yr}^{-1}$. 
{Alt text: This graph shows that the SFR surface density is correlated with the molecular gas mass surface density. 
Plots with higher delta main sequence tend to have higher SFR surface density at the fixed molecular gas mass surface density.}}
\label{fig:resolvedSFR_Mgas}
\end{figure}

Figure \ref{fig:resolvedSFR_Mgas} shows the correlation between the $\Sigma_{\mathrm{SFR}}$ and $\Sigma_{M_{\mathrm{mol}}}$. 
The solid line represents the OLS bisector power-law fit for all points, where the slope and intercept are $1.02 \pm 0.02$ and $-9.01 \pm 0.17$, respectively. 
The dotted line represents a constant SFE of $\Sigma_{\mathrm{SFR}} / \Sigma_{M_{\mathrm{mol}}} = 10^{-9}\ \mathrm{yr}^{-1}$. 
This figure shows a correlation between the molecular gas and star formation, a well-known spatially resolved K--S relation. 
In addition, the figure shows that the K--S relation depends on $\Delta$MS. 
In other words, UMS galaxies tend to be located in the upper end of this relation. 
This weak dependence is consistent with the results of previous studies (\cite{Ellison20b, Morselli20}). 
More details about the spatially resolved K--S relation will be discussed in a companion COMING paper (Takeuchi et al. in prep.).

% figure 3: azimuthally averaged surface density of SFR vs. that of Mmol

\begin{figure}
 \begin{center}
  \includegraphics[width=8cm]{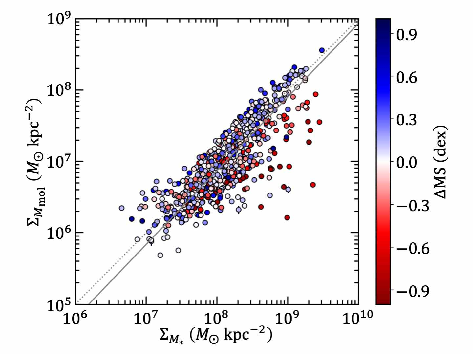}
 \end{center}
\caption{Same as figure \ref{fig:resolvedSFR_Mstar}, except for the relation between averaged molecular gas mass and stellar mass surface density. 
The gray solid line represents the OLS bisector fitting for all data points. 
The gray dotted line represents a constant molecular gas mass fraction of $\Sigma_{M_{\mathrm{mol}}} / \Sigma_{M_{\ast}} = 10^{-1}$. 
{Alt text: This graph shows that the molecular gas mass surface density is correlated with the stellar mass surface density. 
The value of the delta main sequence does not correlate well with the offset from the correlation.}}
\label{fig:resolvedMgas_Mstar}
\end{figure}

Figure \ref{fig:resolvedMgas_Mstar} shows the correlation between the $\Sigma_{M_{\mathrm{mol}}}$ and $\Sigma_{M_{\ast}}$. 
We can observe the relationship of MGMS (\cite{Lin19}). 
The solid line represents the OLS bisector power-law fit for all points, where the slope and intercept are $1.03 \pm 0.02$ and $-1.39 \pm 0.18$, respectively. 
The dotted line represents a constant molecular gas mass to stellar mass ratio of $\Sigma_{M_{\mathrm{mol}}} / \Sigma_{M_{\ast}} = 10^{-1}$. 
This figure shows that the spatially resolved MGMS also weakly depends on $\Delta$MS, and this trend is not sufficiently strong as the azimuthally averaged K--S relation.

% figure 4: azimuthally averaged surface density of Mmol vs. that of M*

\subsection{Radial profiles of sSFR, $f_{\mathrm{gas}}$, and SFE}\label{subsec:radialprof}

Azimuthally averaged $\Sigma_{M_{\ast}}$, $\Sigma_{\mathrm{SFR}}$, and $\Sigma_{M_{\mathrm{mol}}}$ correlate well with each other, although the relations have scatters of $\sim 0.3\,\mathrm{dex}$, as discussed in the previous subsection. 
The radial profiles of sSFR, molecular gas fraction $f_{\mathrm{gas}}$, which is defined as $\Sigma_{M_{\mathrm{mol}}} / \Sigma_{M_{\ast}}$ according to \citet{Sorai19}, and SFE should be investigated.

Figure \ref{fig:radialdist_sSFR} shows the radial profiles of sSFR for the UMS (top left), MS (top right), and LMS (bottom left) galaxies color-coded by $\Delta$MS. 
Averaged values for each $\Delta$MS category are plotted in figure \ref{fig:radialdist_sSFR}d. 
The radial distributions of the sSFRs of UMS and MS galaxies are almost flat, whereas that of LMS galaxies slightly decreases within the $\sim 0.2\,R_{25}$ region. 
In our samples, $0.2\,R_{25}$ approximately corresponds to 3\,kpc. 
As $\Delta$MS decreases, the sSFR consistently decreases at all radii of galaxies: 
the sSFR of UMS galaxies is $\sim 3 \times 10^{-10}\ \mathrm{yr}^{-1}$ on average, whereas that of LMS galaxies is $\lesssim 4 \times 10^{-11}\ \mathrm{yr}^{-1}$. 
MS galaxies exhibit radial profiles that are intermediate between the two groups. 
In other words, the azimuthally averaged sSFR varies with the global sSFR for all radii.

% figure 5: radial distribution of azimuthally averaged sSFR

In the central region within $\lesssim 0.2\,R_{25}$ of LMS galaxies, the azimuthally averaged sSFR decreases toward the galaxy center. 
Abuddro'uf \& Akiyama (2017, 2018) reported that galaxies below the global SFMS, which may be quenched, exhibit a decline in the sSFR in the central region. 
The same results were also obtained for ``green valley'' galaxies, lying below the SFMS but not fully passive (\cite{Belfiore18}). 
\citet{Ellison18} also found that galaxies above the global SFMS have particularly high sSFRs within a central $\sim 3\ \mathrm{kpc}$ region, whereas the sSFR profiles of those below the global SFMS are depressed throughout the galaxy, and that the most significant star formation deficit is observed in the central region. 
Our results are consistent with those of previous studies.

Figure \ref{fig:radialdist_fgas} shows the radial profiles of the $f_{\mathrm{gas}}$. 
We found that $f_{\mathrm{gas}}$ is almost constant in UMS and MS galaxies, similar to the case of the sSFR. 
The $f_{\mathrm{gas}}$ of LMS galaxies is lower than that of UMS or MS galaxies at all radii and especially decreases in the inner region of $\lesssim 0.2\,R_{25}$. 
In the central region, the $f_{\mathrm{gas}}$ of two-thirds of UMS galaxies exceeded 0.1, whereas that of LMS galaxies decreased to $\sim 0.03$ on average. 

% figure 6: radial distribution of azimuthally averaged fgas

We also investigated the radial profiles of the SFE values for UMS, MS, and LMS galaxies, as shown in figure \ref{fig:radialdist_SFE}. 
The SFE is almost constant along the galactocentric radius for each galaxy (\cite{Muraoka19}), and it decreases with radius in LMS galaxies. 
The outermost point of the radial distribution of LMS galaxies comprises a single galaxy; hence, it has little statistical significance. 
In addition, the constant SFE value tends to decrease with decreasing $\Delta$MS. 
In other words, the SFE in LMS galaxies is lower than that in UMS galaxies, in general, as observed in figures \ref{fig:resolvedSFR_Mgas} and \ref{fig:radialdist_SFE}. 
The tendency of a constant radial distribution of the SFE was also observed in the analysis by \citet{Ellison20a} in their moderately star-forming galaxies. 
This result is consistent with that of previous studies from the ALMaQUEST survey, which found that galaxies with low sSFR, such as those in the green valley, exhibit reduced SFE compared to main sequence galaxies (e.g., \cite{Lin22}; \cite{Pan24}). 
Colombo \etal\ (\yearcite{Colombo25a}, \yearcite{Colombo25b}) reported a similar trend that lower $f_{\mathrm{gas}}$ and SFE in LMS galaxies for the EDGE-CALIFA survey samples covered much wider SFR range including galaxies with less star formation activity, such as green valleys and passive galaxies.

% figure 7: radial distribution of azimuthally averaged SFE

\section{Discussion}\label{sec:discussion}

We found that the sSFR, $f_{\mathrm{gas}}$, and SFE in the disk region of galaxies maintain a constant value regardless of their $\Delta$MS, but the constant value decreases with decreasing $\Delta$MS. 
In addition, we observed that galaxies with low (high) global $\Delta$MS tend to decrease (flatten) in their sSFR and $f_{\mathrm{gas}}$ values in the central region of galaxies. 
In this section, we discuss some causes for the above characteristics as well as the exceptional behavior of the radial distribution illustrated in figure \ref{fig:radialdist_fgas} in subsection \ref{subsec:NGC3310}. 

\subsection{Causes of the variation in the radial sSFR profiles}\label{subsec:sSFR_variation}
As shown in figure \ref{fig:radialdist_sSFR}, the sSFR does not change significantly within a galaxy and remains almost constant over the entire galaxy. 
Galaxies with higher global sSFR values tend to have higher azimuthally averaged sSFR values than those with lower global sSFR values.

\begin{figure}
 \begin{center}
  \includegraphics[width=5.38cm]{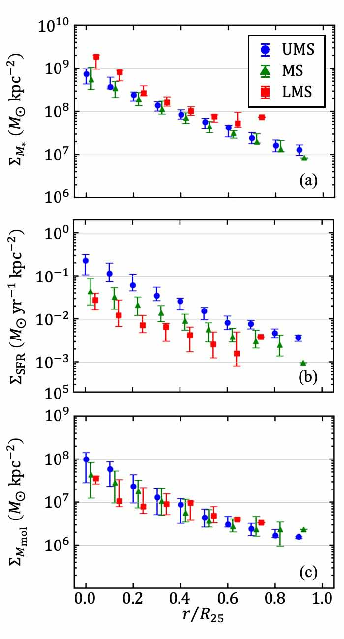}
 \end{center}
\caption{Radial profiles of averaged (a) $\Sigma_{M_{\ast}}$, (b) $\Sigma_{\mathrm{SFR}}$, and (c) $\Sigma_{M_{\mathrm{mol}}}$. 
Blue circles, green triangles, and red squares show the median of $\Sigma_{M_{\ast}}$, $\Sigma_{\mathrm{SFR}}$, or $\Sigma_{M_{\mathrm{mol}}}$ for each radius in each MS group. 
The lower (upper) error bar corresponds to the first (third) quartile value. 
Green triangles are shifted by 0.02 and red squares are shifted by 0.04 to the right from their actual position in $r / R_{25}$. 
{Alt text: The three graphs show the stellar mass surface, SFR surface, and the molecular gas mass surface densities decrease with the galaxy radius. 
LMS galaxies have slightly higher values of the stellar mass surface density than the other two types of galaxies. 
The SFR surface density of UMS galaxies is highest, while that of LMS galaxies is lowest. 
The molecular gas mass surface density of LMS galaxies is slightly lower than those of the other two types of galaxies in the inner region.}}
\label{fig:radialdist_SurfaceDensities}
\end{figure}

Verifying whether this tendency results from a spurious correlation between the stellar mass and stellar mass surface density is necessary. 
Figure \ref{fig:histogram_Mstar} shows the histograms of $\Delta$MS for galaxies with different $M_{\ast}$. 
Higher $\Delta$MS values are observed in galaxies with lower $M_{\ast}$, and more massive galaxies shift to lower $\Delta$MS than less massive galaxies. 
In other words, massive galaxies tend to have lower $\Delta$MS. 
Further, we show the radial distribution of the averaged surface density of $M_{\ast}$, $\Sigma_{M_{\ast}}$ in figure \ref{fig:radialdist_SurfaceDensities}a. 
LMS galaxies also tend to have a slightly higher stellar mass surface density than UMS or MS galaxies. 
However, the radial distribution of the averaged surface density of the SFR, $\Sigma_{\mathrm{SFR}}$, of LMS galaxies also deviates from those of UMS and MS galaxies, as shown in figure \ref{fig:radialdist_SurfaceDensities}b; 
LMS galaxies tend to have lower $\Sigma_{\mathrm{SFR}}$ for all radii. 
Moreover, the shift in $\Sigma_{\mathrm{SFR}}$ exceeds that in $\Sigma_{M_{\ast}}$. 
A systematic error may occur in deriving the stellar mass. 
We used the method of deriving the stellar mass from WISE 3.4\,$\mu$m band. 
\citet{Wen13} showed a different parameter set for early-type galaxies, although these galaxies seem to be redder than the normal spiral galaxies. 
However, if we adopt these ``early-type'' parameter sets for our LMS galaxies, the derived stellar mass and its surface density may become higher than the present values. 
Thus, a systematic difference in stellar mass or stellar mass surface density is not the main cause of the trend that the sSFR surface density, $\Sigma_{\mathrm{sSFR}}$, is lower in all radii in LMS galaxies. 
Many stars have formed, and the star formation activity has declined, resulting in a lower sSFR in LMS galaxies.

% figure 8: histograms of delta MS for different stellar masses

% figure 9: radial distributions of azimuthally averaged Mstar, SFR, & Mmol surface densities

Because the inverse number of the sSFR can be considered as a timescale of star formation, the above-mentioned facts indicate that the timescale in the kpc resolution has not changed locally and that star formation activity has become quiet over the entire galaxy. 
The radial distribution of the averaged surface density of $M_{\mathrm{mol}}$, $\Sigma_{M_{\mathrm{mol}}}$, in figure \ref{fig:radialdist_SurfaceDensities}c does not vary among the UMS, MS, and LMS galaxies in the disk region exceeding $\sim 0.2 R_{25}$. 
Thus, the lower activity of star formation in LMS galaxies is likely caused by a decrease in the SFE.

What causes the SFE decrease over the galaxy disk? 
To examine some global characteristics of galaxies that correlate with $\Delta$MS, we verified the percentage of the Hubble type of galaxies in each group in figure \ref{fig:fraction_HubbleTypes}. 
The fraction of early-type galaxies in LMS galaxies was remarkably higher than those in UMS and MS galaxies. 
Observations of disk galaxies with various bulge fractions (\cite{Kennicutt89}) shows that bulge-dominated galaxies form stars less efficiently than disk-dominated galaxies. 

% figure 10: fraction of Hubble types for delta MS classes

\begin{figure}
 \begin{center}
  \includegraphics[width=8cm]{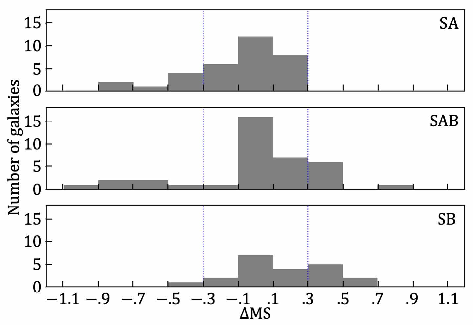}
 \end{center}
\caption{Histograms of $\Delta$MS for different morphological types. 
The vertical dotted lines show $\Delta \mathrm{MS} = \pm0.3$. 
{Alt text: The three histograms exhibit non-barred galaxies tend to have a lower value of the delta main sequence.}}
\label{fig:histogram_bar}
\end{figure}

Figure \ref{fig:histogram_bar} shows the histograms of the $\Delta$MS for different morphological types based on the de Vaucouleurs classification for the sample galaxies (\cite{deVaucouleurs91}). 
Previous studies have shown that star formation activity depends on the presence of a bar structure (e.g., \cite{Abdurrouf18}). 
UMS galaxies are only SB and SAB types, and LMS galaxies are only SA and SAB types, except for one SB galaxy (NGC\,5792). 
MS galaxies comprised SA-, SAB-, and SB-types of galaxies. 
The Kolmogorov--Smirnov (K--S) two-sample test rejects the null hypothesis that SA galaxies in our sample originate from galaxies with higher $\Delta$MS than SAB and SB galaxies, with a $p$-value of 0.018 in the one-sided test. 
This result shows that the existence of the bar may affect the $\Delta$MS in our sample. 
Earlier Hubble types, which expect a larger bulge, may stabilize the gas disk and cause ``morphological quenching'' (\cite{Martig09}; \cite{Quilley22}) in LMS galaxies. 
In contrast, disks in barred spirals become more stable than those in non-barred spirals under the same molecular gas surface density. 
Thus, the above-mentioned tendency that UMS galaxies prefer barred spirals may be caused by the sample selection bias that barred spirals are rather less massive than non-barred spirals in the COMING galaxies.

% figure 11: histograms of delta MS for different bar types

\begin{figure}
 \begin{center}
  \includegraphics[width=8cm]{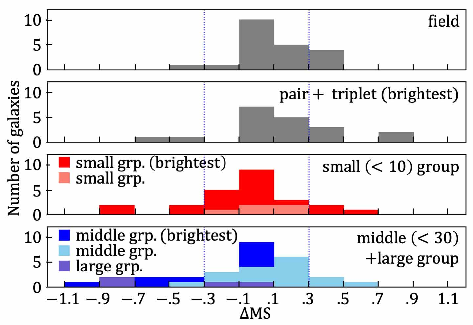}
 \end{center}
\caption{Histograms of $\Delta$MS for different environments. 
The vertical dotted lines represent $\Delta \mathrm{MS} = \pm0.3$. 
{Alt text: The four histograms exhibit field galaxies and galaxies in a pair or triplet system tend to have a higher value of the delta main sequence, while galaxies in a group or cluster tend to have a lower value of the delta main sequence.}}
\label{fig:histogram_group}
\end{figure}

This global depletion of the SFE may be caused by the galaxy environment. 
We classify the sample galaxies into field, pair or triplet system, a small group whose number of galaxies is less than 10, and a middle group whose number of galaxies is less than 30, or a large group that includes the members of the Virgo cluster, respectively, based mainly on \citet{Kourkchi17}, \citet{Garcia93}, and \citet{Giuricin00}. 
The brightest galaxies in each group were also identified. 
Figure \ref{fig:histogram_group} shows the histograms of the $\Delta$MS for different environments. 
Galaxies in larger groups may prefer lower $\Delta$MS. 
The K--S two-sided test rejects the null hypothesis that field and pair/triplet galaxies (galaxies in more sparse environments) and galaxies in larger groups (galaxies in denser environments) originate from the same population, with a $p$-value of 0.0027. 
The one-sided test also rejects the null hypothesis that galaxies in denser environments prefer higher $\Delta$MS than those in sparser environments, with a $p$-value of 0.0013. 

% figure 12: histograms of delta MS for different environment

\begin{figure}
 \begin{center}
  \includegraphics[width=8cm]{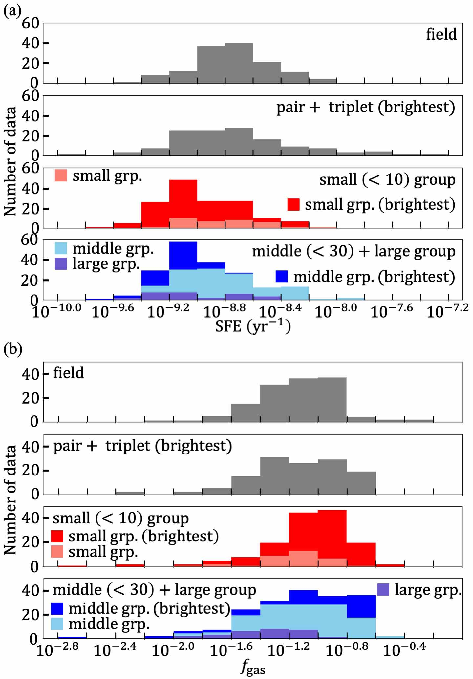}
 \end{center}
\caption{Histograms of (a) SFE and (b) $f_{\mathrm{gas}}$ in azimuthally averaged data for different environments. 
{Alt text: Histograms of SFE show that galaxies in a group or cluster tend to have a lower SFE value compared to galaxies in sparser environments. 
In contrast, the histograms of gas mass fraction show that galaxies in a group or cluster typically exhibit a slightly higher gas mass fraction than their field counterparts.}}
\label{fig:histogram_SFE_fgas_group}
\end{figure}

We also show the histograms of SFE and $f_{\mathrm{gas}}$ in azimuthally averaged data for different environments in figure \ref{fig:histogram_SFE_fgas_group}. 
The K--S one-sided test rejects the null hypothesis that galaxies in denser environments prefer higher SFE than those in sparser environments with, a $p$-value of $3.3 \times 10^{-13}$. 
However, $f_{\mathrm{gas}}$ in galaxies in denser environments exhibit slightly higher values than galaxies in sparser environments. 
The K--S one-sided test rejects the null hypothesis that galaxies in denser environments prefer lower $f_{\mathrm{gas}}$ values than those in sparser environments, with a $p$-value of 0.014. 
Molecular gas accumulates in galaxies in denser environments, at least in our sample, which decreases the SFE in those galaxies and results in less effective star formation. 
\citet{Lisenfeld14} reported a lower SFE, which might be caused by the presence of diffuse or kinematically perturbed gas, was recorded at off-center positions in some galaxies in Hickson Compact Groups. 
Our sample galaxies may also exhibit such an environmental effect on the SFE. 
Further observations tracing dense molecular gases are required to clarify this possibility.

% figure 13: histograms of SFE & fgas for different environment

\subsection{Causes of the variation in the central $f_{\mathrm{gas}}$ across the main sequence}\label{subsec:fgas_variation}
Although the radial profiles of sSFR, $f_{\mathrm{gas}}$, and SFE were roughly flat over all radii based on $\Delta$MS, the sSFR and $f_{\mathrm{gas}}$ decreased in the inner region less than $\sim 0.2 R_{25}$ in LMS galaxies. 
In this subsection, we examine the reasons for these decreasing trends.

First, the decrease in $f_{\mathrm{gas}}$ in the central region of LMS galaxies was not caused by the adoption of a constant $X_{\mathrm{CO}}$. 
\citet{Yasuda23} showed that the $X_{\mathrm{CO}}$ of the COMING galaxies generally increases radially outward, although the number of investigated galaxies is lower than one-fourth of our number of samples. 
If we adopt the variable $X_{\mathrm{CO}}$, $f_{\mathrm{gas}}$ is expected to increase steeply with increasing radius.

\begin{figure}
 \begin{center}
  \includegraphics[width=8cm]{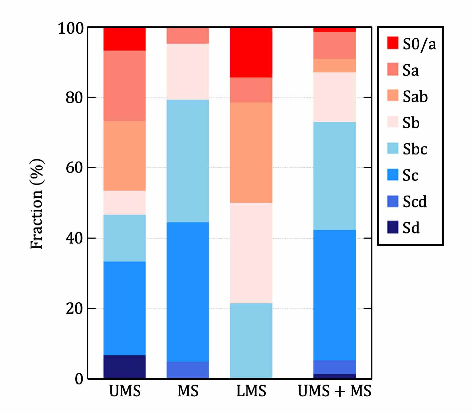}
 \end{center}
\caption{Fraction of the Hubble types of sample galaxies in each main sequence group. 
{Alt text: This bar fraction graph shows that the fraction of early-type galaxies is higher for LMS galaxies, while the fraction of late-type galaxies is higher for MS galaxies. 
Although the fraction of early- and late-type galaxies compete for UMS galaxies, 
the fraction of late-type galaxies is much higher than that of LMS galaxies.}}
\label{fig:fraction_HubbleTypes}
\end{figure}

Considering that the radial distribution of $\Sigma_{\mathrm{SFR}}$ increases inwardly, whereas $\Sigma_{M_{\mathrm{mol}}}$ has a rather flat radial distribution in the inner region, and that the radial distribution of SFE is flat and rather higher than that in the region of $\gtrsim 0.2 R_{25}$, the decrease in sSFR at the central region can be caused mainly by the $f_{\mathrm{gas}}$ decrease. 
A possible reason for the central $f_{\mathrm{gas}}$ decrease is the consumption of molecular gas by star formation. 
As shown in figure \ref{fig:fraction_HubbleTypes}, the Hubble type of LMS galaxies tends to form earlier than that of UMS or MS galaxies. 
As the Hubble type is an indicator of the bulge size of the samples, and a more dominant bulge in earlier disk galaxies comprises numerous stars, a considerable amount of molecular gas is expected to be consumed in the central region (\cite{Young82}).

The ejection of molecular gas from the galactic disk by the active galactic nucleus (AGN) and stellar feedback is another possibility. 
One of the commonly known scenarios for reducing or removing the molecular gas, particularly in the central region of galaxies, is AGN feedback. 
The AGN feedback heats up or expels the surrounding gas and prevents galaxies from subsequent star formation. 
Some local galaxies with high far-infrared surface brightness attain extremely high molecular gas outflow rates of $\gtrsim 10\,M_{\solar}\ \mathrm{yr}^{-1} - 1000\,M_{\solar}\ \mathrm{yr}^{-1}$, and the outflow mass is up to $10 \%$ of the molecular gas mass of the galaxy (\cite{Lutz20}).  
This scenario is supported by previous observations of the low gas fraction in AGN host galaxies (e.g., \cite{Maiolino12}) and AGN-driven molecular gas outflow (e.g., \cite{Cicone14}; \cite{Feruglio15}). 
In addition, AGN host galaxies lie in the ``green valley" or below the SFMS (e.g., \cite{Lin17}). 
This result shows that the AGN accounts for the transition from the star-forming phase to the quiescent phase in the star formation history of the galaxy. 
A trend of lower SFE and $f_{\mathrm{gas}}$ values in AGN hosts was also reported by \citet{Bazzi25}.
As for stellar feedback, multiple supernovae resulting from starbursts account for the molecular gas outflow from the central region of the galaxy (e.g., \cite{Nakai87}, \cite{Salak20}). 
Therefore, we verified whether our samples were Seyfert or low-ionization nuclear emission-line region (LINER) galaxies, which are AGN-type galaxies, as well as H \emissiontype{II}-region-like or starburst galaxies. 
The percentages of AGN host galaxies in UMS, MS, and LMS galaxies were $27\%$, $27\%$, and $71\%$, respectively. 
However, the fractions of H \emissiontype{II}-region-like or starburst galaxies in UMS, MS, and LMS galaxies were $73 \%$, $71 \%$, and $36 \%$, respectively. 

\begin{figure}
 \begin{center}
  \includegraphics[width=8cm]{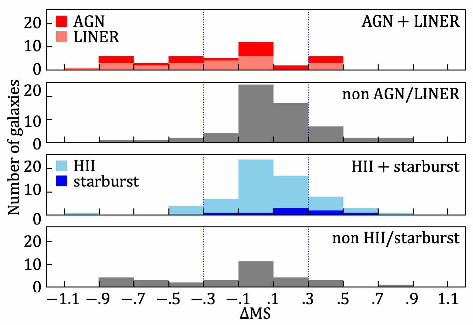}
 \end{center}
\caption{Histograms of $\Delta$MS for nuclear activity types. 
The vertical dotted lines show $\Delta \mathrm{MS} = \pm0.3$. 
{Alt text: Histograms of the delta main sequence show that galaxies with the AGN tend to have a lower value of the delta main sequence, and actively star-forming galaxies tend to have a higher value of the delta main sequence.}}
\label{fig:histogram_activity}
\end{figure}

These trends are observed in figure \ref{fig:histogram_activity}, which shows the histograms of $\Delta$MS for galaxies with and without AGN, as well as those with and without an actively star-forming nucleus. 
The K--S two-sample test rejects the null hypothesis that AGN and non-AGN samples originate from the same population, with a $p$-value of $4.3 \times 10^{-5}$ in the two-sided test, and also the null hypothesis that galaxies with AGN have a higher $\Delta$MS than non-active galaxies, with a $p$-value of $2.1 \times 10^{-5}$ in the one-sided test. 
The test rejects the null hypothesis that star-forming and other galaxies originate from the same population, with a $p$-value of 0.0016 in the two-sided test, and also the null hypothesis that star-forming galaxies have lower $\Delta$MS than other galaxies, with a $p$-value of $8.1 \times 10^{-4}$. 
This is consistent with the AGN feedback framework, which states that the presence of the AGN reduces the available cold gas in the central region.

% figure 14: histograms of delta MS for different activity types

\begin{figure}
 \begin{center}
  \includegraphics[width=8cm]{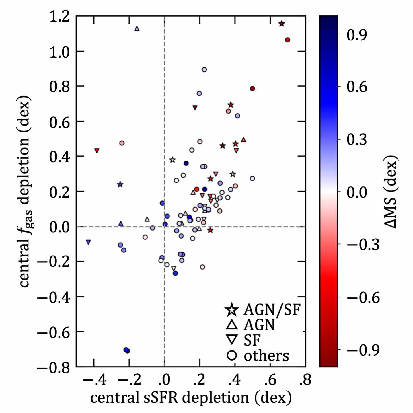}
 \end{center}
\caption{Correlation of central depletion of $f_{\mathrm{gas}}$ and that of sSFR. 
The colors correspond to individual $\Delta$MS of galaxies. 
Marks represent the following. Star: AGN with H \emissiontype{II}-region-like or starburst nucleus, upward triangle: AGN without enhanced star formation activity, downward triangle: H \emissiontype{II}-region-like or starburst nucleus, and circle: other galaxies. 
{Alt text: The central depletion of the gas mass fraction weakly correlates with the central depletion of the sSFR.}}
\label{fig:depletion_fgas_sSFR}
\end{figure}

We define the central depletion of $f_{\mathrm{gas}}$ and sSFR as follows: 
\begin{eqnarray}
\mathrm{central\ depletion\ of}\ f_{\mathrm{gas}} \ \mathrm{(dex)} = \nonumber \\
\left< \mathrm{log}_{10}\ f_{\mathrm{gas}}\left( 0.3 \leq r / R_{25} \leq 0.5 \right) \right> \nonumber \\
- \left< \mathrm{log}_{10}\ f_{\mathrm{gas}}\left( r / R_{25} < 0.2 \right) \right>, 
\end{eqnarray}
\begin{eqnarray}
\mathrm{central\ depletion\ of\ sSFR} = \nonumber \\
\left< \mathrm{log}_{10}\ \mathrm{sSFR}\left( 0.3 \leq r / R_{25} \leq 0.5 \right) \right> \nonumber \\
- \left< \mathrm{log}_{10}\ \mathrm{sSFR}\left( r / R_{25} < 0.2 \right) \right>, 
\end{eqnarray}
where the average represents the geometric mean. 
Figure \ref{fig:depletion_fgas_sSFR} shows the correlation between these two depletion quantities with color-coded $\Delta$MS. 
Both depletions correlate with a Pearson's product moment correlation coefficient of 0.49 and a Spearman's rank correlation coefficient of 0.55. 
The gradient of the correlation is higher than unity, indicating that the central depletion of $f_{\mathrm{gas}}$ is a cause of the central depletion of the sSFR rather than the opposite cause and effect. 
LMS galaxies prefer both central depletions. 
Moreover, most galaxies with central enhancement (i.e., negative depletion) exhibit no AGN activity. 
However, the scatter in the plot and different marks that intermingle should be considered, indicating that multiple mechanisms cause $\Delta$MS and the central depletion.

% figure 15: central depletions of fgas & sSFR

Finally, we investigated the plausibility of the lack of a molecular gas supply from the disk region to the central region. 
\citet{Brownson20} obtained the same result as ours: $f_{\mathrm{gas}}$ and SFE are suppressed at all radii, and $f_{\mathrm{gas}}$ is reduced by $\sim 1\, \mathrm{dex}$ in the central regions of their green valley sample galaxies. 
They estimated that the reduction in $f_{\mathrm{gas}}$ was caused by a decrease in the gas supply rather than an ejection of molecular gas. 
As shown in figure \ref{fig:fraction_HubbleTypes}, LMS galaxies shift to the earlier Hubble type. 
The gaseous viscosity in the galaxy disk may cause different transfer rates of molecular gas toward the inner region (\cite{Courteau96}; \cite{Zhang00}). 
An effective gas transport toward the central region led to the development of the bulge; hence, a lower $f_{\mathrm{gas}}$ is expected in early-type galaxies.

The central $f_{\mathrm{gas}}$ variation may also be related to the existence of a bar structure. 
The bar potential removes the angular momentum of the molecular gas rotating around the center of the galaxy, causing the molecular gas to fall into the central region (e.g., \cite{Downes96}). 
Previous studies have demonstrated that barred galaxies have higher molecular gas concentrations in the central region than non-barred galaxies (e.g, \cite{Sakamoto99}; \cite{Jogee05}; \cite{Sheth05}; \cite{Kuno07}). 
Thus, we expect barred galaxies to have a high $f_{\mathrm{gas}}$ in the inner region, where molecular gas is supplied from the outer region through the bar, which sustains continuous active star formation in the central region. 
In contrast, non-barred galaxies have no such effective channels for supplying molecular gas to the inner region. 
Therefore, molecular gas is consumed during star formation, and $f_{\mathrm{gas}}$ (and subsequently SFR) decreases. 
However, we have to take notice that this discussion implicitly assumes a closed gas system; a more accurate treatment is beyond the scope of this study.

Our suggestion is consistent with those of previous studies on UMS galaxies; the SFR is higher in the central region of barred galaxies than in non-barred galaxies (e.g., \cite{Hawarden86}; \cite{Devereux87}; \cite{Puxley88}; \cite{Ho97}; \cite{Ellison11}; \cite{Oh12}; \cite{Zhou15}; \cite{Hogarth24}). 
\citet{Chown19} showed that the degree of enhancement of the star formation activity in the central region of galaxies is positively correlated with the molecular gas concentration for barred galaxies. 
Therefore, bar structures may be essential in enhancing star formation activity in the central region of UMS galaxies in our sample. 
However, barred galaxies in the COMING samples are biased toward actively star-forming galaxies (\cite{Sorai19}).

The origin of the decrease in the mass of the molecular gas remains controversial and may not be due to a single process. 
However, the central depletion of $f_{\mathrm{gas}}$ and sSFR may support the inside--out quenching of star formation; galaxies tend to quench their star formation activities from the central region (\cite{Perez13}). 
The evidence of inside--out quenching has also been reported in previous studies (e.g., \cite{Tacchella15}; \cite{GonzalezDelgado16}; \cite{Belfiore18}; \cite{Tacchella18}; \cite{Ellison18}; \cite{Abdurrouf17}; \cite{Abdurrouf18}). 
To determine which scenario is most plausible, additional data are required. 
These data include spatially resolved metallicity, which shows the history of star formation within the galaxy (e.g., \cite{Sanchez12}); molecular gas distribution perpendicular to the galactic plane, which shows the existence of extraplanar molecular gas (e.g., \cite{Combes13}); and galactic structure, which indicates the effect of characteristic gas motion.

\subsection{What determines global $\Delta$MS}
\label{subsec:UMS_LMS}
The preceding subsections summarize the properties of UMS and LMS galaxies as follows: 
\begin{itemize}
\item UMS galaxies tend to be less massive and/or barred spiral galaxies.
Their $f_{\mathrm{gas}}$ values are similar to those of MS galaxies, whereas their SFE values are higher than those of MS galaxies, particularly in the central region.
\item LMS galaxies tend to be massive and/or hosts for AGN.  
Their $f_{\mathrm{gas}}$ values are lower than those of MS galaxies, particularly in the central region. 
Conversely, their SFE values are comparable to those of MS galaxies, although they are lower in the outer region.
\end{itemize}
\noindent
These findings show that a higher SFE causes galaxies to have higher $\Delta$MS values, and a lower $f_{\mathrm{gas}}$ value causes them to have lower $\Delta$MS values. 
Disk instability tends to be considered in galaxies with less prominent stellar concentrations (or bulges), leading to the formation of stellar structures such as bars or spiral arms. 
The molecular gas in these systems is frequently compressed, and the SFE increases across the entire disk. 
Such galaxies have been observed as UMS galaxies. 
Conversely, for LMS galaxies, a considerable time, which is indicated by an sSFR of $\sim 3 \times 10^{-11}\,\mathrm{yr}^{-1}$, has passed since their major star-forming epoch; therefore, molecular gas has been substantially consumed throughout the galaxy. 
The outer atomic gas (the precursor to the molecular gas) of galaxies in groups may be stripped, and the gas supply may be truncated by tidal interactions. 
In addition, a larger bulge in LMS galaxies decreases the sSFR and $f_{\mathrm{gas}}$ in the central region. 
Our sample galaxies may have evolved in this manner, however, comprehensive data, particularly from more widespread samples, are required for definitive confirmation.

\subsection{Peculiar radial behavior of NGC\,3310}
\label{subsec:NGC3310}
As shown in figure \ref{fig:radialdist_fgas}a, NGC\,3310, which is the highest $\Delta$MS galaxy in our sample, exhibits an extremely low $f_{\mathrm{gas}}$ ($\sim 3\%$) in the inner region and a monotonous increase toward the outer radius. 
The two possible explanations for this finding are as follows. 
The first is the rapid consumption of molecular gas. 
NGC\,3310 is a starburst galaxy probably triggered by a recent minor interaction (\cite{Miralles-Caballero14}). 
Because starburst galaxies have a high SFR, stars form rapidly, consuming a considerable amount of molecular gas. 
Therefore, an extremely low $f_{\mathrm{gas}}$ was observed in NGC\,3310 compared with the other UMS galaxies. 
The second is that the metal content is low and the $X_{\mathrm{CO}}$ factor is high, 
although the latter may be implausible for actively star-forming galaxies (e.g., \cite{Bolatto13}). 
\citet{Muraoka19} reported that NGC\,3310 exhibits a moderately low metallicity ($0.2 - 0.4\,Z_{\solar}$; \cite{Pastoriza93}), which indicates a higher $X_{\mathrm{CO}}$ than the standard value of the Milky Way (\cite{Bolatto13}), and thus an underestimation of $\Sigma_{\mathrm{gas}}$, and finally a lower $f_{\mathrm{gas}}$.

\section{Conclusions}\label{sec:conclusions}
In this study, we investigated the relationship between the surface density of the SFR, stellar mass, and molecular gas mass on the $\sim 1\,\mathrm{kpc}$ scale for 92 galaxies from the COMING project. 
We categorized these galaxies into three groups based on the deviation of the global SFR from the main sequence for the samples ($\Delta$MS) to link the variation in the azimuthally averaged SFMS with the global SFR. 
The three groups were divided as $\Delta\mathrm{MS} > 0.3\,\mathrm{dex}$ (UMS), $-0.3 \leq \Delta\mathrm{MS} \leq 0.3$ (MS), and $\Delta\mathrm{MS} < -0.3$ (LMS). 
Using the maps of the $\Sigma_{M_{\ast}}$, $\Sigma_{\mathrm{SFR}}$, and $\Sigma_{M_{\mathrm{mol}}}$, we obtained these surface densities averaged over concentric tilted rings with $0.1\,R_{25}$ radius. 
The results and implications of this study are as follows: 
\begin{itemize}
\item The azimuthally averaged SFR and stellar mass surface densities ($\Sigma_{\mathrm{SFR}} - \Sigma_{M_{\ast}}$) relation exhibited good correlation, similar to the global SFR and $M_{\ast}$ relation. 
In addition, this relation showed that lower $\Delta$MS galaxies suppress their $\Sigma_{\mathrm{SFR}}$ over the galaxy.
The radial distributions of the sSFR, $f_{\mathrm{gas}}$, and SFE were almost constant over the entire galaxy, particularly in UMS and MS galaxies. 
The values decreased with the $\Delta$MS. 
The systematic decrease in the sSFR with the $\Delta$MS may be attributed to a lower SFE rather than a higher $\Sigma_{M_{\ast}}$.
\item We found that the fraction of early-type disk galaxies was higher in LMS galaxies than in UMS and MS galaxies and that all UMS galaxies had a strong or weak bar, whereas LMS galaxies had a weak or no bar. 
Moreover, group galaxies preferred LMS galaxies. 
This shows that the Hubble type, that is, the difference in bulge size, bar, and galaxy environment, may stabilize the molecular gas in the disk of LMS galaxies and cause a lower SFE in the entire disk.
\item LMS galaxies in our samples showed that sSFR decreases in the central region of less than $\sim 0.2 R_{25}$, whereas UMS galaxies sustain a high sSFR in the same region. 
The radial profiles of $f_{\mathrm{gas}}$ also exhibited the same trend as those of the sSFR. 
LMS galaxies showed extremely low $f_{\mathrm{gas}} \lesssim \mathrm{0.05}$, while UMS galaxies showed high $f_{\mathrm{gas}} \gtrsim 0.1$ in the central region. 
The central drop of the sSFR in the LMS galaxies may have resulted from $f_{\mathrm{gas}}$, considering that $\Sigma_{\mathrm{SFR}}$ increases inward and $\Sigma_{M_{\mathrm{mol}}}$ reaches the ceiling in the central region.
\item UMS galaxies tended to be less massive and/or barred spirals, while LMS galaxies tended to be massive and/or host the AGN. 
UMS galaxies had $f_{\mathrm{gas}}$ values similar to those of MS galaxies, whereas LMS galaxies had lower $f_{\mathrm{gas}}$ values than MS galaxies, particularly in the central region. 
UMS galaxies had a higher SFE than MS galaxies, particularly in the central region. 
In contrast, LMS galaxies showed SFE comparable to that of MS galaxies overall, but lower SFE in their outer regions.
\item These results show that the increase in SFE primarily increased the $\Delta$MS of a galaxy, whereas a decrease in $f_{\mathrm{gas}}$ caused a decrease in the $\Delta$MS. 
UMS galaxies tended to have less centrally concentrated stellar distributions, which may easily induce spiral arms and bar structures. 
This, in turn, leads to the frequent compression of molecular gas, thereby increasing the SFE. 
However, primary star formation in LMS galaxies occurred in the past, probably consuming their molecular gas supply. 
Additionally, external factors such as tidal forces may have stripped away the atomic gas that forms molecular gas, leading to a reduction in the molecular gas content.
\item The consumption of molecular gas during star formation in the past accounted for the decrease in $f_{\mathrm{gas}}$ in the central region of LMS galaxies. 
This scenario is consistent with the fact that LMS galaxies tend to have an earlier Hubble type than UMS or MS galaxies.
\item Similarly, the AGN feedback may also account for the decrease in $f_{\mathrm{gas}}$ in the central region. 
In our sample, the fraction of galaxies with AGN or LINER was higher in  LMS galaxies than in UMS or MS galaxies, whereas the fraction of galaxies with H \emissiontype{II}-region-like or starburst nuclei was higher in UMS and MS galaxies than in LMS galaxies.
\item The lack of a molecular gas supply also results in the $f_{\mathrm{gas}}$ in the central region of LMS galaxies. 
Galaxies without strong bar structures cannot effectively supply molecular gas to the central region. 
Molecular gas transfer through the gaseous viscosity may also induce gas depletion in the central region.
\item The sSFR, $f_{\mathrm{gas}}$, and SFE at the disk region show constant values regardless of $\Delta$MS, but these values decreased with decreasing $\Delta$MS. 
This indicates that lower $\Delta$MS galaxies tend to move toward the end of their star formation across the galaxy, and the depression of star formation is caused by the decreasing SFE. 
Combined with the fact that star formation was suppressed in the central region of LMS galaxies, which was mainly caused by the decrease in $f_{\mathrm{gas}}$, we observed that star formation ends in the galactic center. 
This indicates an inside--out quenching scenario.
\end{itemize}

%%%%%%%%%%%%%%%%%%%%%%%%%%%%%%%%%%%%%%% 

% table1: galaxy parameters
%\begin{landscape}
%\centering
\begin{longtable}{lccccllllccc}
\caption{Parameters of the sample galaxies.}
\label{tab:galpars} \\ 
\hline
Galaxy & Distance & PA & $i$ & $R_{25}$ & \multicolumn{4}{c}{Galaxy properties} & $\mathrm{log}_{10} M_{\ast}$ & $\mathrm{log}_{10} \mathrm{SFR}$ & $\Delta$MS \\
 & (Mpc) & $(^{\circ})$ & $(^{\circ})$ & $(\timeform{''})$ & & & & & $(M_{\solar})$ & $(M_{\solar}\,\mathrm{yr}^{-1})$ & (dex) \\
 (1) & (2) & (3) & (4) & (5) & (6) & (7) & (8) & (9) & (10) & (11) & (12) \\
\hline
\endfirsthead
\hline
Galaxy & Distance & PA & $i$ & $R_{25}$ & \multicolumn{4}{c}{Galaxy properties} & $\mathrm{log}_{10} M_{\ast}$ & $\mathrm{log}_{10} \mathrm{SFR}$ & $\Delta$MS \\
 & (Mpc) & $(^{\circ})$ & $(^{\circ})$ & $(\timeform{''})$ & & & & & $(M_{\solar})$ & $(M_{\solar}\,\mathrm{yr}^{-1})$ & (dex) \\
 (1) & (2) & (3) & (4) & (5) & (6) & (7) & (8) & (9) & (10) & (11) & (12) \\
\hline
\endhead
\hline
\endfoot
\hline
\multicolumn{12}{@{}l@{}}{\hbox to 0pt{\parbox{160mm}{\footnotesize
Notes. 
\par\noindent
Columns: (1) Galaxy name. (2) Distance of galaxies in Mpc. (3) Position angle of the major axis. (4) Inclination of galaxy disks. (5) Optical radius ($R_{25}$) in arcsec. (6) Morphology for bar. (7) Nuclear activity. ``AL'' represents the galaxy is identified as AGN or LINER. (8) Nuclear star formation activity. ``SF'' represents the galaxy has H \emissiontype{II}-region-like or starburst nucleus. (9) Galaxy environment. ``FL'': field galaxy, ``BPT'': the brightest galaxy in a pair or triplet system, ``BSG'': the brightest galaxy in a small-member group, ``MSG'': a member in a small-member group, ``BMG'': the brightest galaxy in a middle-member group, ``MMG'': a member in a middle-member group, and ``MLG'': a member in a large-member group or the Virgo cluster. (10) Logarithmic stellar mass within $R_{25}$ in $M_{\solar}$. (11) Logarithmic SFR within $R_{25}$ in $M_{\solar}\,\mathrm{yr}^{-1}$. (12) $\Delta$MS in dex. (2) -- (5) are cited or calculated from \citet{Sorai19}. 
}\hss}}
\endlastfoot

NGC\,0150 & 24.20 & 108.9 & 59.3 & 116.7 & SB && SF & BPT & 10.42 & 0.69 & 0.24 \\
NGC\,0157&12.10 &-136.0 &48.0 &125.1 &SAB&&SF&FL&10.15 &0.23 &0.06 \\
NGC\,0337&18.90 &119.6&44.5&86.4 &SB&&SF&FL&9.94 &0.26 &0.31 \\
NGC\,0470&41.30 &155.4&58.0 &84.6 &SA&&SF&MMG&10.72 &0.90 &0.14 \\
NGC\,0613&26.4&-54.1&38.8&165.0 &SB&AL&SF&BPT&11.08 &1.16 &0.02 \\
NGC\,0628&9.020 &20&7&314.1 &SA&&SF&BPT&10.25 &0.32 &0.05 \\
NGC\,0660&13.60 &-138.5&72&249.6 &SB&AL&SF&BPT&10.42 &0.79 &0.34 \\
NGC\,0701&20.80 &44.7&58.6&73.5 &SB&&SF&BPT&10.00 &0.16 &0.15 \\
NGC\,1022&18.5&85.0 &30&72.0 &SB&&SF&MMG&10.14 &0.69 &0.54 \\
NGC\,1084&20.90 &-141.6&57.2&97.2 &SA&&SF&BSG&10.55 &0.85 &0.27 \\
NGC\,1087&14.90 &1.4&50.5&111.6 &SAB&&SF&MSG&10.00 &0.25 &0.24 \\
NGC\,1241&61.40 &-45&52&84.6 &SB&AL&&MSG&11.12 &1.06 &-0.12 \\
UGC\,2765&19.5&150.0 &57&119.4 &&&&BPT&10.21 &0.98 &0.75 \\
IC\,0356&21.6&105.0 &43&157.5 &SA&&&BMG&11.23 &0.71 &-0.58 \\
NGC\,1530&20.40 &-172&45&137.1 &SB&&&FL&10.40 &0.45 &0.03 \\
NGC\,2146&27.7&-43.5&62&180.9 &SB&&SF&FL&11.27 &1.79 &0.46 \\
NGC\,2273&31.6&51&53&97.2 &SB&AL&&BPT&10.65 &0.82 &0.13 \\
NGC\,2268&30.60 &-112&58&97.2 &SAB&&&MMG&10.60 &0.56 &-0.08 \\
NGC\,2276&36.8&-113&48&84.6 &SAB&&&MMG&10.73 &1.08 &0.30 \\
NGC\,2633&29.10 &-176.3&50.1&73.5 &SB&&SF&BSG&10.39 &0.94 &0.52 \\
NGC\,2681&16.40 &116.6&11.2&108.9 &SAB&AL&&BPT&10.41 &-0.11 &-0.54 \\
NGC\,2742&27.8&-93.5&59.9&90.6 &SA&&&MMG&10.40 &0.29 &-0.13 \\
NGC\,2715&17.50 &-159.1&67.8&147.0 &SAB&&&MSG&10.00 &-0.03 &-0.05 \\
NGC\,2775&17.0 &163.5&35.4&128.1 &SA&&&BSG&10.67 &-0.19 &-0.89 \\
NGC\,2748&19.80 &-138.8&72.8&90.6 &SA&&SF&BPT&10.10 &0.17 &0.06 \\
NGC\,2782&15.7&-105&30&104.1 &SAB&AL&SF&BSG&9.91 &0.23 &0.32 \\
NGC\,2841&14.60 &152.6&73.7&243.9 &SA&AL&&BSG&10.85 &0.12 &-0.77 \\
NGC\,2903&9.46&-155&67&377.7 &SAB&&SF&BSG&10.60 &0.59 &-0.05 \\
NGC\,2967&22.2&64.0 &16.5&90.6 &SA&&&BPT&10.20 &0.30 &0.08 \\
NGC\,2976&3.63&-25.5&64.5&176.7 &SA&&SF&MLG&9.10 &-0.88 &0.05 \\
NGC\,3166&22.0 &-100.4&55.7&143.7 &SAB&AL&SF&BMG&10.77 &-0.10 &-0.91 \\
NGC\,3169&23.20 &-123.7&39.0 &131.1 &SA&AL&SF&MMG&10.86 &0.43 &-0.47 \\
NGC\,3147&39.30 &142.79&35.19&116.7 &SA&AL&SF&BSG&11.26 &1.00 &-0.32 \\
NGC\,3198&13.40 &-145.0 &71.5&255.3 &SB&&SF&FL&10.11 &0.15 &0.03 \\
NGC\,3310&11.6&150&56&92.7 &SAB&&SF&BPT&9.79 &0.59 &0.80 \\
NGC\,3338&23.70 &97.1&60.9&176.7 &SA&&&BSG&10.45 &0.36 &-0.13 \\
NGC\,3344&9.82&-37.3&27.0 &212.4 &SAB&&SF&FL&10.08 &0.03 &-0.06 \\
NGC\,3351&10.7&-168&41&222.3 &SB&&SF&MLG&10.38 &0.25 &-0.16 \\
NGC\,3367&30.8&51&30&75.3 &SB&AL&&BSG&10.49 &0.86 &0.34 \\
NGC\,3359&20.80 &-8&51&217.2 &SB&&SF&FL&10.30 &0.41 &0.09 \\
NGC\,3368&9.900 &169.0 &57.5&227.7 &SAB&AL&&MLG&10.49 &-0.34 &-0.85 \\
NGC\,3370&25.60 &-38.1&55.1&94.8 &SA&&&FL&10.09 &0.27 &0.16 \\
NGC\,3437&25.10 &-61.5&65.9&75.3 &SAB&&SF&FL&10.27 &0.53 &0.24 \\
NGC\,3521&14.20 &-19&63&328.8 &SAB&AL&SF&BSG&11.05 &0.82 &-0.28 \\
NGC\,3583&27.9&131.3&38.6&84.6 &SB&&&BSG&10.54 &0.50 &-0.07 \\
NGC\,3627&9.04&176&52&273.6 &SAB&AL&&BMG&10.59 &0.56 &-0.07 \\
NGC\,3655&38.40 &-100.3&23.5&46.5 &SA&&SF&BPT&10.67 &0.79 &0.08 \\
NGC\,3672&27.40 &7.8&67.2&125.1 &SA&&SF&BSG&10.62 &0.63 &-0.02 \\
NGC\,3675&19.60 &176&67.8&176.7 &SA&&SF&FL&10.84 &0.44 &-0.45 \\
NGC\,3686&15.9&19.5&35.2&97.2 &SB&&&BSG&9.99 &0.00 &-0.01 \\
NGC\,3810&16.40 &-154.3&42.2&128.1 &SA&&SF&BSG&10.32 &0.44 &0.10 \\
NGC\,3813&23.30 &83.1&68.2&67.2 &SA&&&BPT&10.20 &0.33 &0.11 \\
NGC\,3888&39.7&121.2&41.8&52.2 &SAB&&SF&BPT&10.51 &0.61 &0.06 \\
NGC\,3893&15.7&-13&30&134.1 &SAB&&SF&MMG&10.28 &0.40 &0.10 \\
NGC\,3938&17.9&-154.0 &20.9&161.1 &SA&&SF&MMG&10.43 &0.47 &0.01 \\
NGC\,3949&19.10 &-58.2&52.9&86.4 &SA&&SF&MMG&10.17 &0.37 &0.18 \\
NGC\,4030&29.90 &29.6&39.0 &125.1 &SA&&SF&BSG&11.06 &1.02 &-0.10 \\
NGC\,4041&30.2&-138.7&23.4&80.7 &SA&&SF&MMG&10.67 &0.89 &0.18 \\
NGC\,4045&33.70 &-92.1&48.4&80.7 &SAB&AL&SF&FL&10.58 &0.62 &0.01 \\
NGC\,4085&20.80 &-104.7&75.0 &84.6 &SAB&&SF&MMG&9.99 &0.05 &0.05 \\
NGC\,4088&14.50 &-126.8&68.9&172.5 &SAB&&SF&MMG&10.38 &0.55 &0.14 \\
NGC\,4258&7.31&-29&72&558.6 &SAB&AL&&BMG&10.52 &0.03 &-0.52 \\
NGC\,4303&16.5&-36.4&27.0 &193.8 &SAB&AL&SF&BMG&10.77 &0.89 &0.08 \\
NGC\,4433&44.10 &5&64&65.7 &SAB&&SF&BPT&10.62 &1.09 &0.44 \\
NGC\,4527&16.5&69.5&70&185.1 &SAB&AL&SF&BMG&10.68 &0.66 &-0.06 \\
NGC\,4536&16.5&-54.5&64.2&227.7 &SAB&&SF&MMG&10.46 &0.73 &0.24 \\
NGC\,4559&7.31&-36.8&63.1&321.6 &SAB&&SF&FL&9.69 &-0.19 &0.13 \\
NGC\,4579&16.5&92.1&41.7&176.7 &SAB&AL&&MLG&10.81 &0.15 &-0.71 \\
NGC\,4605&5.55&-67&69&172.5 &SB&&SF&FL&9.37 &-0.61 &0.03 \\
NGC\,4602&37.80 &100.7&67.9&101.7 &SAB&AL&&MSG&10.74 &0.69 &-0.09 \\
NGC\,4632&14.40 &60.5&65.9&92.7 &SA&&SF&MSG&9.68 &-0.19 &0.13 \\
NGC\,4666&14.70 &-135&70&137.1 &SAB&AL&SF&BSG&10.56 &0.58 &-0.01 \\
NGC\,4750&26.1&-50.0 &40&61.2 &SA&AL&&BPT&10.56 &0.24 &-0.36 \\
NGC\,4818&11.90 &-175.5&67.2&128.1 &SAB&&SF&FL&9.92 &0.40 &0.47 \\
NGC\,5005&18.00 &67.0 &66.7&172.5 &SAB&AL&&BMG&10.91 &0.50 &-0.46 \\
NGC\,5055&9.04&98&61&377.7 &SA&AL&SF&BSG&10.71 &0.48 &-0.27 \\
NGC\,5248&13.00 &103.9&38.6&185.1 &SAB&AL&SF&BMG&10.36 &0.42 &0.03 \\
NGC\,5364&18.2&-144.4&47.9&202.8 &SA&&SF&MMG&10.43 &0.25 &-0.22 \\
NGC\,5678&35.70 &-177.5&56.9&99.3 &SAB&AL&&BPT&10.85 &0.86 &-0.04 \\
NGC\,5665&18.20 &154.7&51.7&57.3 &SAB&&SF&BPT&9.76 &0.11 &0.36 \\
NGC\,5676&34.70 &-131.9&59.8&119.4 &SA&&SF&BMG&10.93 &0.97 &-0.01 \\
NGC\,5713&19.5&-157&33&82.5 &SAB&&SF&MMG&10.28 &0.69 &0.38 \\
NGC\,5792&26.40 &-98.5&64&207.6 &SB&&SF&BSG&10.92 &0.64 &-0.32 \\
NGC\,6015&19.00 &-150&62&161.1 &SA&&SF&FL&10.22 &0.21 &-0.03 \\
NGC\,6503&6.25&-60.2&73.5&212.4 &SA&AL&SF&FL&9.66 &-0.59 &-0.24 \\
NGC\,6574&41.90 &165&45&42.3 &SAB&AL&&FL&10.95 &1.13 &0.14 \\
NGC\,6951&23.30 &135&30&116.7 &SAB&AL&&FL&10.85 &0.85 &-0.04 \\
NGC\,7331&13.90 &167.7&75.8&314.1 &SA&AL&SF&BSG&10.91 &0.67 &-0.29 \\
NGC\,7479&36.80 &-158&51&122.1 &SB&AL&&FL&11.01 &1.39 &0.33 \\
NGC\,7625&23.0 &-151.4&37.4&47.4 &SA&&SF&BPT&10.17 &0.36 &0.17 \\
NGC\,7721&28.00 &-164.2&69.8&106.5 &SA&&&FL&10.40 &0.36 &-0.06 \\
NGC\,7798&32.6&70.5&31.9&41.4 &SB&&SF&FL&10.27 &0.49 &0.19 \\
\end{longtable}
%\end{landscape}

%%%%%%%%%%%%%%%%%%%%%%%%%%%%%%%%%%%%%%%

%\section*{Supplementary data} 

\begin{ack}
%Acknowledgement should be placed at end of main text.
%(NOT after the Appendix.)
The authors thank the referee for valuable suggestions that have helped improve this manuscript. 
This work is based on one of the legacy programs of the Nobeyama 45\,m radio telescope, which is operated by Nobeyama Radio Observatory, a branch of National Astronomical Observatory of Japan. 
This research has made use of the NASA/IPAC Extragalactic Database (NED) which is operated by the Jet Propulsion Laboratory, California Institute of Technology, under contract with the National Aeronautics and Space Administration. 
This publication makes use of data products from the Wide-field Infrared Survey Explorer, which is a joint project of the University of California, Los Angeles, and the Jet Propulsion Laboratory/California Institute of Technology, funded by the National Aeronautics and Space Administration. 
This work has been supported in part by the Japan Society for the Promotion of Science (JSPS) Grants-in-Aid for Scientific Research (21H01128 and 24H00247), and the Collaboration Funding of the Institute of Statistical Mathematics ``Machine-Learning-Based Cosmogony: From Structure Formation to Galaxy Evolution''. 
We would like to thank Editage (www.editage.jp) for English language editing.
\end{ack}

%\appendix %%%%%%%%%%%%%%%%%%%%%%%%%%%%%%%%%%%%%%%%%%%%%%%%%%%%%%%%
%\section*{Case of single paragraph}
% No section number is necessary. Add ``*'' after \verb/\section/.

%%%% 
%\section{Case of two or more paragraphs}

% Text of appendix

%\section{Case of two or more paragraphs}

% Text of appendix

% Any journal's BST file (e.g., apj.bst) can be used as PASJ's BST is unavailable.    
% \bibliographystyle{****}
% \bibliography{****}

\begin{thebibliography}{}
\bibitem[Abdurro'uf \& Akiyama(2018)]{Abdurrouf18}
  Abdurro'uf, \& Akiyama, M. \ 2018, \mnras, 479, 5083
\bibitem[Abdurro'uf \& Akiyama(2017)]{Abdurrouf17}
  Abdurro'uf, \& Akiyama, M. \ 2017, \mnras, 469, 2806
\bibitem[Bazzi et al.(2025)]{Bazzi25}
  Bazzi, Z., Colombo, D., Bigiel, F., Kalinova, V., Villanueva, V., Sanchez, S.~F., Bolatto, A.~D., \& Wong, T. \ 2025, \aap, 697, A149
\bibitem[Belfiore et al.(2018)]{Belfiore18}
  Belfiore, F., \etal \ 2018, \mnras, 477, 3014
\bibitem[Bolatto et al.(2017)]{Bolatto17}
  Bolatto, A.~D., \etal \ 2017, \apj, 846, 159
\bibitem[Bolatto et al.(2013)]{Bolatto13}
  Bolatto, A.~D., Wolfire, M., \& Leroy, A.~K. \ 2013, \araa, 51, 207
\bibitem[Brownson et al.(2020)]{Brownson20}
  Brownson, S., Belfiore, F., Maiolino, R., Lin, L., \& Carniani, S. \ 2020, \mnras, 498, L66
\bibitem[Bundy et al.(2015)]{Bundy15}
  Bundy, K., \etal \ 2015, \apj, 798, 7
\bibitem[Cano-D\'{i}az et al.(2016)]{Cano-Diaz16}
  Cano-D\'{i}az, M., \etal \ 2016, \apjl, 821, L26
\bibitem[Casasola et al.(2017)]{Casasola17}
  Casasola, V., \etal \ 2017, \aap, 605, A18
\bibitem[Chown et al.(2019)]{Chown19}
  Chown, R., \etal \ 2019, \mnras, 484, 5192
\bibitem[Cicone et al.(2014)]{Cicone14}
  Cicone, C., \etal \ 2014, \aap, 562, A21
\bibitem[Colombo et al.(2025a)]{Colombo25a}
  Colombo, D., \etal \ 2025a, \aap, 699, A366
\bibitem[Colombo et al.(2025b)]{Colombo25b}
  Colombo, D., \etal \ 2025b, \aap, 699, A367
\bibitem[Combes et al.(2013)]{Combes13}
  Combes, F., Garc\'{i}a-Burillo, S., Braine, J., Schinnerer, E., Walter, F., \& Colina, L. \ 2013, \aap, 550, A41
\bibitem[Courteau et al.(1996)]{Courteau96}
  Courteau, S., de Jong, R.~S., \& Broeils, A.~H. \ 1996, \apjl, 457, L73
\bibitem[Daddi et al.(2007)]{Daddi07}
  Daddi, E., \etal \ 2007, \apj, 670, 156
\bibitem[Davis \& Seaquist(1983)]{Davis83}
  Davis, L.~E., \& Seaquist, E.~R. \ 1983, \apjs, 53, 269
\bibitem[de Vaucouleurs et al.(1991)]{deVaucouleurs91}
  de Vaucouleurs, G., de Vaucouleurs, A., Corwin, H.~G., Jr., Buta, R.~J., Paturel, G., \& Fouque, P. \ 1991, Third Reference Catalogue of Bright Galaxies (New York: Springer)
\bibitem[Devereux (1987)]{Devereux87}
  Devereux, N. \ 1987, \apj, 323, 91
\bibitem[Downes et al.(1996)]{Downes96}
  Downes, D., Reynaud, D., Solomon, P.~M., \& Radford, S.~J.~E. \ 1996, \apj, 461, 186
\bibitem[Elbaz et al.(2018)]{Elbaz18}
  Elbaz, D., \etal \ 2018, \aap, 616, A110
\bibitem[Elbaz et al.(2007)]{Elbaz07}
  Elbaz, D., \etal \ 2007, \aap, 468, 33
\bibitem[Ellison et al.(2024)]{Ellison24}
  Ellison, S.~L., \etal \ 2024, \mnras, 527, 10201
\bibitem[Ellison et al.(2020b)]{Ellison20b}
  Ellison, S.~L., \etal \ 2020b, \mnras, 493, L39
\bibitem[Ellison et al.(2011)]{Ellison11}
  Ellison, S.~L., Nair, P., Patton, D.~R., Scudder, J.~M., Mendel, J.~T., \& Simard, L. \ 2011, \mnras, 416, 2182
\bibitem[Ellison et al.(2018)]{Ellison18}
  Ellison, S.~L., S\'{a}nchez, S.~F., Ibarra-Medel, H., Antonio, B., Mendel, J.~T., \& Barrera-Ballesteros, J. \ 2018, \mnras, 474, 2039
\bibitem[Ellison et al.(2020a)]{Ellison20a}
  Ellison, S.~L., Thorp, M.~D., Pan, H.~-A., Lin, L., Scudder, J.~M., Bluck, A.~F.~L., S\'{a}nchez, S.~F., \& Sargent, M. \ 2020a, \mnras, 492, 6027
\bibitem[Enia et al.(2020)]{Enia20}
  Enia, A., \etal \ 2020, \mnras, 493, 4107
\bibitem[Feruglio et al.(2015)]{Feruglio15}
  Feruglio, C., \etal \ 2015, \aap, 583, A99
\bibitem[Garcia (1993)]{Garcia93}
  Garcia, A.~M. \ 1993, \aaps, 100, 47
\bibitem[Genzel et al.(2015)]{Genzel15}
  Genzel, R., \etal \ 2015, \apj, 800, 20
\bibitem[Giuricin et al.(2000)]{Giuricin00}
  Giuricin, G., Marinoni, C., Ceriani, L., \& Pisani, A. \ 2000, \apj, 543, 178
\bibitem[Gonz\'{a}lez Delgado et al.(2016)]{GonzalezDelgado16}
  Gonz\'{a}lez Delgado, R.~M., \etal \ 2016, \aap, 590, A44
\bibitem[Hawarden et al.(1986)]{Hawarden86}
  Hawarden, T.~G., Mountain, C.~M., Leggett, S.~K., \& Puxley, P.~J. \ 1986, \mnras, 221, 41P
\bibitem[Ho et al.(1997)]{Ho97}
  Ho, L.~C., Filippenko, A.~V., \& Sargent, W.~L.~W. \ 1997, \apj, 487, 568
\bibitem[Hogarth et al.(2024)]{Hogarth24}
  Hogarth, L.~M., Saintonge, A., Davis, T.~A., Ellison, S.~L., Lin, L., L\'{o}pez-Cob\'{a}, C., Pan, H.~-A., \& Thorp, M.~D. \ 2024, \mnras, 528, 6768
\bibitem[Hsieh et al.(2017)]{Hsieh17}
  Hsieh, B.~C., \etal \ 2017, \apjl, 851, L24
\bibitem[Isobe et al.(1990)]{Isobe90}
  Isobe, T., Feigelson, E.~D., Akritas, M.~G., \& Babu, G.~J. \ 1990, \apj, 364, 104
\bibitem[Jogee et al.(2005)]{Jogee05}
  Jogee, S., Scoville, N., \& Kenney, J.~D.~P. \ 2005, \apj, 630, 837
\bibitem[Kennicutt(1989)]{Kennicutt89}
  Kennicutt, R.~C., Jr. \ 1989, \apj, 344, 685
\bibitem[Kennicutt \& Evans(2012)]{Kennicutt12}
  Kennicutt, R.~C., Jr., \& Evans, N.~J., II \ 2012, \araa, 50, 531
\bibitem[Koopmann \& Kenney(2004)]{Koopmann04}
  Koopmann, R.~A., \& Kenney, J.~D.~P. \ 2004, \apj, 613, 866
\bibitem[Kourkchi \& Tully(2017)]{Kourkchi17}
  Kourkchi, E., \& Tully, R.~B. \ 2017, \apj, 843, 16
\bibitem[Kuno et al.(2007)]{Kuno07}
  Kuno, N., \etal \ 2007, \pasj, 59, 117
\bibitem[Lin et al.(2022)]{Lin22}
  Lin, L., \etal \ 2022, \apj, 926, 175
\bibitem[Lin et al.(2020)]{Lin20}
  Lin, L., \etal \ 2020, \apj, 903, 145
\bibitem[Lin et al.(2019)]{Lin19}
  Lin, L., \etal \ 2019, \apjl, 884, L33
\bibitem[Lin et al.(2017)]{Lin17}
  Lin, L., \etal \ 2017, \apj, 851, 18
\bibitem[Lisenfeld et al.(2014)]{Lisenfeld14}
  Lisenfeld, U., Appleton, P.~N., Cluver, M.~E., Guillard, P., Alatalo, K., \& Ogle, P. \ 2014, \aap, 570, A24
\bibitem[Lutz et al.(2020)]{Lutz20}
  Lutz, D., \etal \ 2020, \aap, 633, A134
\bibitem[Maiolino et al.(2012)]{Maiolino12}
  Maiolino, R., \etal \ 2012, \mnras, 425, L66
\bibitem[Martig et al.(2009)]{Martig09}
  Martig, M., Bournaud, F., Teyssier, R., \& Dekel, A. \ 2009, \apj, 707, 250
\bibitem[Martin et al.(2005)]{Martin05}
  Martin, D.~C., \etal \ 2005, \apjl, 619, L1
\bibitem[Matharu et al.(2022)]{Matharu22}
  Matharu, J., \etal \ 2022, \apj, 937, 16
\bibitem[Miralles-Caballero et al.(2014)]{Miralles-Caballero14}
  Miralles-Caballero, D., Rosales-Ortega, F.~F., D\'{i}az, A.~I., Ot\'{i}-Floranes, H., P\'{e}rez-Montero, E., \& S\'{a}nchez, S.~F. \ 2014, \mnras, 445, 3803
\bibitem[Morselli et al.(2020)]{Morselli20}
  Morselli, L., \etal \ 2020, \mnras, 496, 4606
\bibitem[Muraoka et al.(2019)]{Muraoka19}
  Muraoka, K., \etal \ 2019, \pasj, 71, S15
\bibitem[Nakai et al.(1987)]{Nakai87}
  Nakai, N., Hayashi, M., Handa, T., Sofue, Y., \& Hasegawa, T. \ 1987, \pasj, 39, 685
\bibitem[Noeske et al.(2007)]{Noeske07}
  Noeske, K.~G., \etal \ 2007, \apjl, 660, L43
\bibitem[Oh et al.(2012)]{Oh12}
  Oh, S., Oh, K., \& Yi, S.~K. \ 2012, \apjs, 198, 4
\bibitem[Pan et al.(2024)]{Pan24}
  Pan, H.~-A., \etal \ 2024, \apj, 964, 120
\bibitem[Pastoriza et al.(1993)]{Pastoriza93}
  Pastoriza, M.~G., Dottori, H.~A., Terlevich, E., Terlevich, R., \& D\'{i}az, A.~I. \ 1993, \mnras, 260, 177
\bibitem[P\'{e}rez et al.(2013)]{Perez13}
  P\'{e}rez, E., \etal \ 2013, \apjl, 764, L1
\bibitem[Puxley et al.(1988)]{Puxley88}
  Puxley, P.~J., Hawarden, T.~G., \& Mountain, C.~M. \ 1988, \mnras, 231, 465
\bibitem[Quilley \& de Lapparent(2022)]{Quilley22}
  Quilley, L., \& de Lapparent, V. \ 2022, \aap, 666, A170
\bibitem[Saintonge et al.(2016)]{Saintonge16}
  Saintonge, A., \etal \ 2016, \mnras, 462, 1749
\bibitem[Saintonge et al.(2012)]{Saintonge12}
  Saintonge, A., \etal \ 2012, \apj, 758, 73
\bibitem[Sakamoto et al.(1999)]{Sakamoto99}
  Sakamoto, K., Okumura, S.~K., Ishizuki, S., \& Scoville, N.~Z. \ 1999, \apjs, 124, 403
\bibitem[Salak et al.(2020)]{Salak20}
  Salak, D., Nakai, N., Sorai, K., \& Miyamoto, Y. \ 2020, \apj, 901, 151
\bibitem[S\'{a}nchez et al.(2012)]{Sanchez12}
  S\'{a}nchez, S.~F., \etal \ 2012, \aap, 538, A8
\bibitem[Sargent et al.(2014)]{Sargent14}
  Sargent, M.~T., \etal \ 2014, \apj, 793, 19
\bibitem[Schmidt(1959)]{Schmidt59}
  Schmidt, M.\ 1959, \apj, 129, 243
\bibitem[Scoville et al.(2016)]{Scoville16}
  Scoville, N., \etal \ 2016, \apj, 820, 83
\bibitem[Sheth et al.(2005)]{Sheth05}
  Sheth, K., Vogel, S.~N., Regan, M.~W., Thornley, M.~D., \& Teuben, P.~J. \ 2005, \apj, 632, 217
\bibitem[Silverman et al.(2018)]{Silverman18}
  Silverman, J.~D., \etal \ 2018, \apj, 867, 92
\bibitem[Silverman et al.(2015)]{Silverman15}
  Silverman, J.~D., \etal \ 2015, \apjl, 812, L23
\bibitem[Sorai et al.(2019)]{Sorai19}
  Sorai, K., \etal \ 2019, \pasj, 71, S14
\bibitem[Speagle et al.(2014)]{Speagle14}
  Speagle, J.~S., Steinhardt, C.~L., Capak, P.~L., \& Silverman, J.~D. \ 2014, \apjs, 214, 15
\bibitem[Tacchella et al.(2018)]{Tacchella18}
  Tacchella, S., \etal \ 2018, \apj, 859, 56
\bibitem[Tacchella et al.(2015)]{Tacchella15}
  Tacchella, S., \etal \ 2015, Science, 348, 314
\bibitem[Tacconi et al.(2018)]{Tacconi18}
  Tacconi, L.~J., \etal \ 2018, \apj, 853, 179
\bibitem[Utomo et al.(2017)]{Utomo17}
  Utomo, D., \etal \ 2017, \apj, 849, 26
\bibitem[Vorontsov-Velyaminov et al.(2001)]{Vorontsov-Velyaminov01}
  Vorontsov-Velyaminov, B.~A., Noskova, R.~I., \& Arkhipova, V.~P. \ 2001, Astronomical \& Astrophysical Transactions, 20, 717
\bibitem[Wen et al.(2013)]{Wen13}
  Wen, X.~-Q., Wu, H., Zhu, Y.~-N., Lam, M.~I., Wu, C.~-J., Wicker, J., \& Zhao, Y.~-H. 2013, \mnras, 433, 2946
\bibitem[Wright et al.(2010)]{Wright10}
  Wright, E.~L., \etal \ 2010, \aj, 140, 1868
\bibitem[Wuyts et al.(2013)]{Wuyts13}
  Wuyts, S., \etal \ 2013, \apj, 779, 135
\bibitem[Wylezalek et al.(2022)]{Wylezalek22}
  Wylezalek, D., \etal \ 2022, \mnras, 510, 3119
\bibitem[Yasuda et al.(2023)]{Yasuda23}
  Yasuda, A., \etal \ 2023, \pasj, 75, 743
\bibitem[Young \& Scoville(1982)]{Young82}
  Young, J.~S., \& Scoville, N. \ 1982, \apjl, 260, L41
\bibitem[Zhang \& Wyse(2000)]{Zhang00}
  Zhang, B., \& Wyse, R.~F.~G. \ 2000, \mnras, 313, 310
\bibitem[Zhou et al.(2015)]{Zhou15}
  Zhou, Z.~-M., Cao, C., \& Wu, H. \ 2015, \aj, 149, 1
\end{thebibliography}

% figure 1

% figure 2

% figure 3

% figure 4

% figure 5
\begin{onecolumn}
\begin{figure}
 \begin{center}
  \includegraphics[width=12cm]{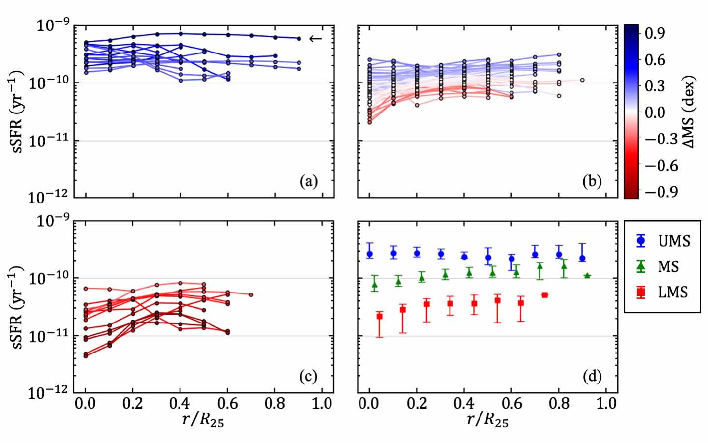}
 \end{center}
\caption{Radial profiles of sSFR on (a) UMS, (b) MS, and (c) LMS galaxies. 
The colors correspond to individual $\Delta$MS of galaxies. 
(d) Comparison of the radial profiles of the three groups. 
Blue circles, green triangles, and red squares represent the median of sSFR for each radius in each MS group. 
The lower (upper) error bar corresponds to the first (third) quartile value. 
For easy demonstration, green triangles are shifted by 0.02 and red squares are shifted by 0.04 to the right from the actual position in $r / R_{25}$. 
The data indicated by the arrow in (a) is NGC\,3310. 
{Alt text: These graphs show that the sSFR is roughly constant across the galaxy radius, and the constant values decrease with the delta main sequence. 
The sSFR tends to decrease in the inner region within 0.2 times the optical radius in LMS galaxies.}}
\label{fig:radialdist_sSFR}
\end{figure}
\end{onecolumn}

% figure 6
\begin{figure}
 \begin{center}
  \includegraphics[width=12cm]{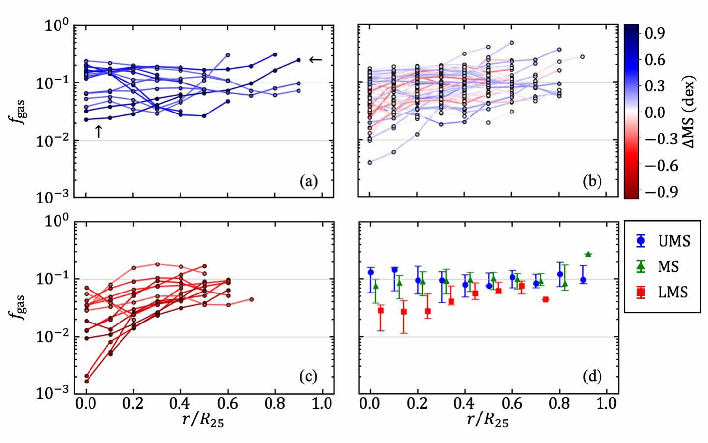}
 \end{center}
\caption{Same as figure \ref{fig:radialdist_sSFR} but for $f_{\mathrm{gas}}$. 
{Alt text: These graphs show that the gas mass fraction is roughly constant across the galaxy radius, and the constant values decrease with the delta main sequence. 
The gas mass fraction tends to decrease in the inner region within 0.2 times the optical radius in LMS galaxies.}}
\label{fig:radialdist_fgas}
\end{figure}

% figure 7
\begin{figure}
 \begin{center}
  \includegraphics[width=12cm]{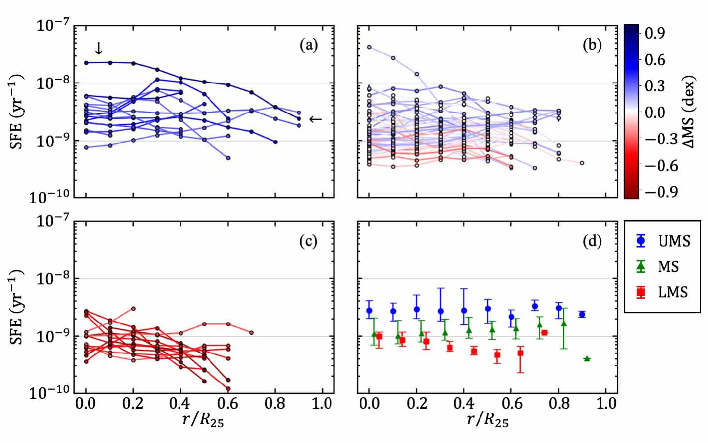}
 \end{center}
\caption{Same as figure \ref{fig:radialdist_sSFR} but for SFE. 
{Alt text: These graphs show that the SFE is roughly constant across the galaxy radius, except for LMS galaxies. 
The values of SFE tend to decrease with the delta main sequence. 
The SFE tends to decrease with radius in LMS galaxies.}}
\label{fig:radialdist_SFE}
\end{figure}

% figure 8
\begin{twocolumn}
\begin{figure}
 \begin{center}
  \includegraphics[width=8cm]{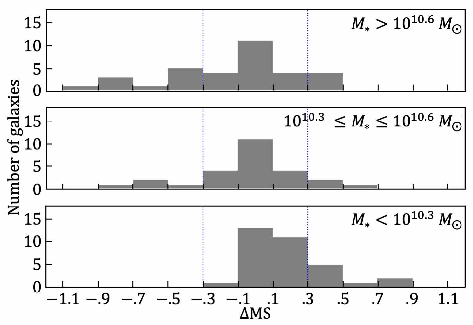}
 \end{center}
\caption{Histograms of $\Delta$MS for different $M_{\ast}$. 
The vertical dotted lines show $\Delta \mathrm{MS} = \pm0.3$. 
{Alt text: The three histograms exhibit massive galaxies tend to have a lower value of the delta main sequence.}}
\label{fig:histogram_Mstar}
\end{figure}
\end{twocolumn}

% figure 9

\clearpage
% figure 10

% figure 11

% figure 12

% figure 13

% figure 14

% figure 15

\end{document}